%% file: arXiv_submission/paperCFD1.tex
\pdfoutput=1
\RequirePackage{atlaslatexpath}

\documentclass[UKenglish,texlive=2016,nonatbib]{elsarticle}

\usepackage[centering, scale=0.75]{geometry}

\DeclareUnicodeCharacter{2212}{\textminus}

\usepackage[backend=bibtex]{atlaspackage}
\usepackage{atlasbiblatex}
\usepackage{setspace}
\usepackage{atlasphysics}

\usepackage{multirow}
\usepackage{bigstrut}
\usepackage{adjustbox}

\usepackage{siunitx}

\graphicspath{{./}}

\usepackage{paperCFD1-defs}

\hypersetup{pdftitle={ATLAS document},pdfauthor={The ATLAS Collaboration}}

\begin{document}

\title{Characterisation of the Thermoflow due to the Dry Nitrogen Flushing Scheme in the ATLAS Inner Tracker using Computational Fluid Dynamics}


\author[up]{Muaaz Bhamjee}
\author[uj]{Matthew Connell\corref{cor1}}
\ead{matthew.peter.connell@cern.ch}
\author[uj]{Simon Connell}
\author[uj]{Emmanuel Igumbor}
\author[up]{Lerothodi Leeuw}
\author[ukzn]{Pedro Mafa}
\author[slac]{Marco Oriunno}
\author[nikhef]{Marcel Vreeswijk}

\cortext[cor1]{Corresponding author}

\address[up]{Department of Mechanical and Aeronautical Engineering, University of Pretoria, Lynnwood Road, Hatfield, Pretoria 0028, South Africa}
\address[uj]{Department of Mechanical Engineering Science, University of Johannesburg, Corner of Kingsway and University Road, Auckland Park, Johannesburg 2006, South Africa}
\address[ukzn]{University of KwaZulu-Natal, Astrophysics Research Centre, Discipline of Mathematics, School of Agriculture and Science, Private Bag X54001, Durban 4000, South Africa}
\address[slac]{SLAC National Accelerator Laboratory, 2575 Sand Hill Road, Menlo Park, California 94025, United States of America}
\address[nikhef]{Nikhef, National Institute for Subatomic Physics, Science Park 105, 1098 XG Amsterdam, Netherlands}


\begin{abstract}
    The planned High Luminosity upgrade to the Large Hadron Collider at CERN aims to increase the instantaneous luminosity peak to about $7.5 \times 10^{34} \mbox{cm}^{-2}\mbox{s}^{- 1}$. The ATLAS detector will be extensively re-designed to meet the challenges of this upgrade. This paper focuses on the use of computational fluid dynamics to characterise the thermoflow in order to model the dry nitrogen flushing scheme in the Common Environmental Monitoring and Interlock System for the ATLAS Inner Tracker as part of the upgrade process.
    The Technical Design Report considers the possibility for the bi-phase CO$_2$ coolant temperature to drop to as low as −55$^{\circ}$C in the case of a fault. The specification for the highest Relative Humidity within the ITk volume is therefore equivalent to a dew point temperature at or below −60$^{\circ}$C in order to prevent condensation which could damage the detector electronics. The design accommodates for humidity monitoring to detect the onset of such events and dry nitrogen flushing to remove moisture.
    Therefore, it is important to thoroughly understand all consequences of atmospheric air ingress due to air-leaks and/or air-ingress from the outlets due to the over-pressure. The computational fluid dynamics model presented in this study was used to provide quantitative and qualitative insight into the various operational and failure conditions, informing engineering design changes to optimise the flushing scheme and ensure that the ITk remains dry and within the design specification of the acceptable dew point range.
\end{abstract}



\maketitle


\section{Introduction}
\label{sec:Intro}
\input{1_introduction}


\section{Computational Fluid Dynamics Model}
\label{sec:Theory}
\input{2_Math_CFD}




%

\section{Results}
\label{sec:Results}
\input{3_results}
\FloatBarrier

\section{Conclusions}
\label{sec:Discussion}
\input{4_discussion}


\FloatBarrier


\section*{Acknowledgements}

\input{Acknowledgements}



\printbibliography



\end{document}

%% file: 1_introduction.tex
The ATLAS Detector~\cite{PERF-2007-01, ATLAS:2019tdj} at the Large Hadron Collider (LHC) in CERN, Geneva, Switzerland is engaged in an upgrade process in preparation for the planned High Luminosity LHC (HL-LHC)~\cite{2020-HL-LHC-TDR, ATLAS-TDR-25, Phase_2}, which aims to achieve a total integrated luminosity of 3000 fb$^{-1}$, increasing the amount of data the detectors record by a factor of 10 and the collision rate by a factor of 3.75.
This is predicted to lead to a proportional 10-fold increase in the integrated radiation dose received by detector equipment, necessitating improvements and new technologies in order to maintain detector performance standards. As part of the upgrade process, the ATLAS Inner Detector (ID) will be completely replaced by a new detector, the Inner Tracker (ITk), to allow more precise tracking at higher collision rates while withstanding the harsh radiation environment~\cite{ATLAS-TDR-25, ATLAS-TDR-30}.

The ITk is an all-silicon, active-element detector responsible for measuring the trajectories of charged particles in the region closest to the collision point. 
The ATLAS detector uses a right-handed coordinate system with the origin at the collision point, the $z$-axis pointing along the beamline and the $y$-axis pointing upwards, in the opposite direction to gravity. Conceptually, the ITk geometry can be simplified to a hollow cylinder parallel to the $z$-axis, containing layers of smaller detector cylinders and layers of detector discs perpendicular to the $z$-axis.

\begin{figure}[!t]
  \begin{center}
    {\includegraphics[width=0.9\textwidth]{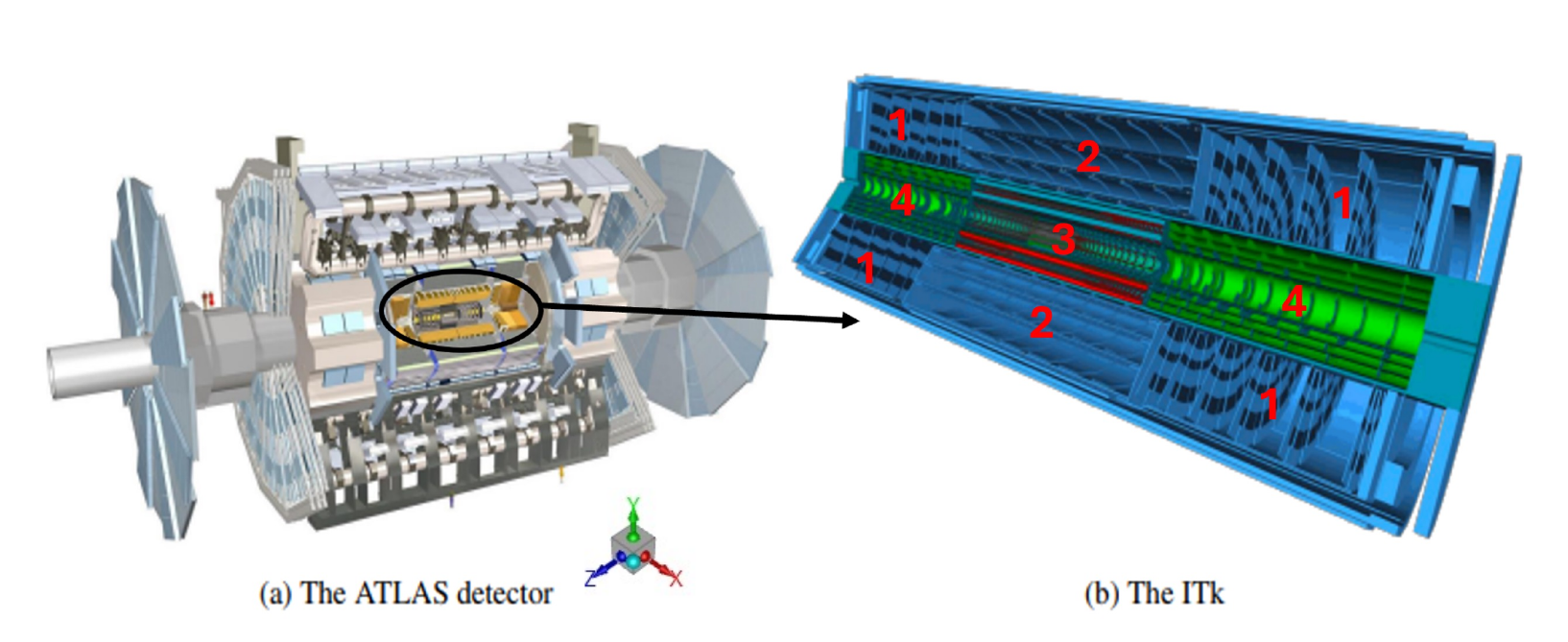}}
    \caption{(a) The ATLAS detector, with the circled region corresponding to (b) the ITk~\cite{ATLAS-TDR-25}. The numbers indicate different regions: 1 Strips Endcaps, 2 Strips Barrel, 3 Pixels Barrel and 4 Pixels Endcaps regions.
      \label{fig:ITk1}}
  \end{center}
\end{figure}

\begin{figure}[!t]
    \begin{center}
        \subfloat[\label{fig:ITk2-octant}]
            {\includegraphics[width=0.54\textwidth]{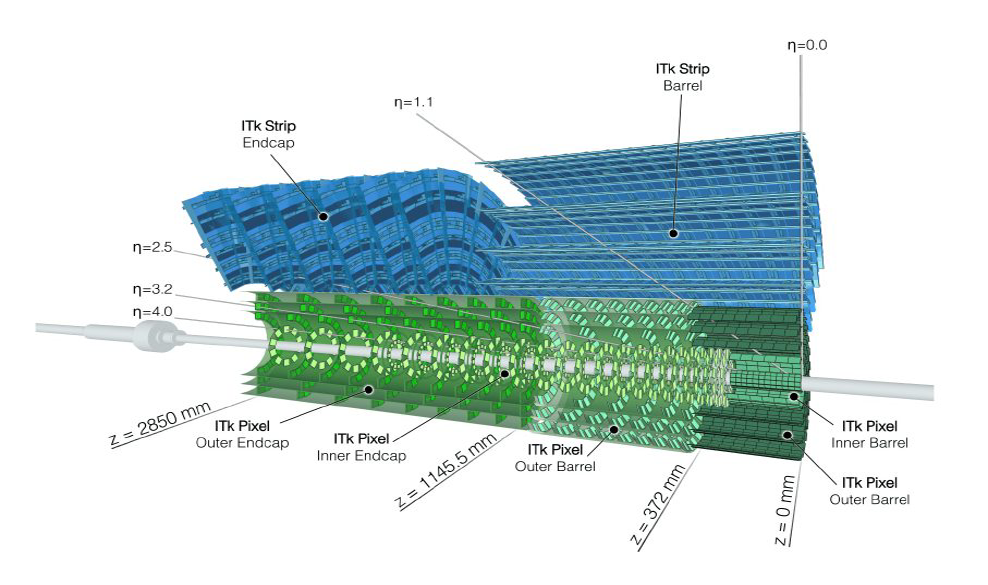}}
        \subfloat[\label{fig:ITk2-rz-plane}]
            {\includegraphics[width=0.42\textwidth]{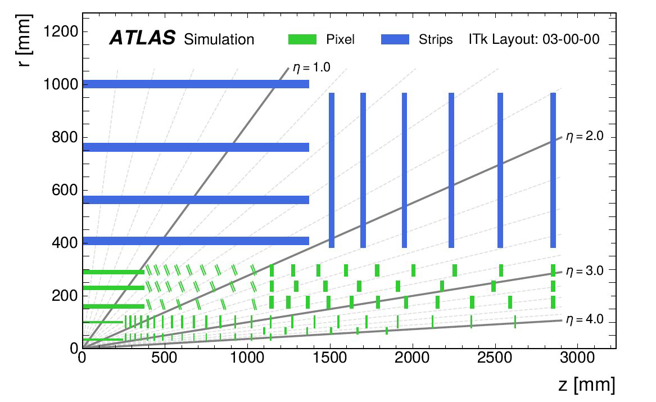}}
        \caption{An octant of the ITk in different views (a) in 3D with different parts labelled and detector discs visible. (b) a cross-section in the $rz$-plane~\cite{ATLAS-TDR-25}.
        \label{fig:ITk2}}
    \end{center}
\end{figure}

As shown in Figures~\ref{fig:ITk1} and ~\ref{fig:ITk2}, the ITk is divided into Pixels and Strips sections based on its detector elements (silicon Pixels and Strips). The silicon Pixels are deployed close to the interaction point, where the highest granularity is required, while silicon Strips are used at larger radii. Both Pixels and Strips sections are configured into central Barrel regions flanked by two symmetrical Endcap regions, as shown in Figure~\ref{fig:ITk1}(b) and Figure~\ref{fig:ITk2-octant}. The Strip detector has four Barrel and six Endcap layers, as shown in Figure~\ref{fig:ITk2-rz-plane}, with active areas grouped in rectangular staves for the Barrel region and truncated triangular petals for the Endcap region, geometrically ensuring that particle trajectories cross the sensors as perpendicularly as possible.

\begin{figure}[!ht]
    \begin{center}
        \includegraphics[width=0.7\linewidth]{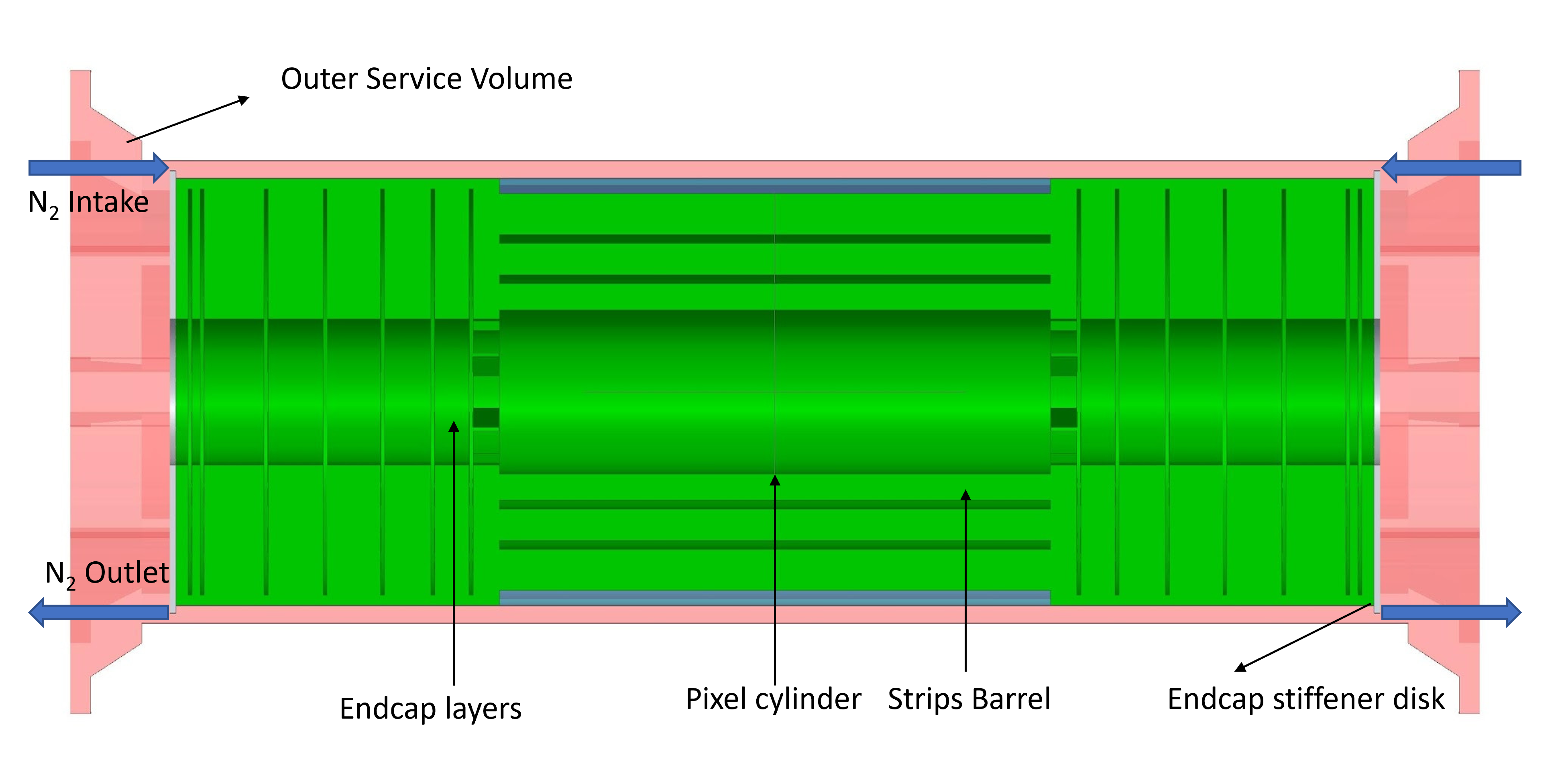}
        \caption{A cross-section of the simplified ITk volume in the $x=0$ plane, with inlets and outlets for the N$_2$ flushing scheme labelled.
        \label{fig:Strips-OSV_XS}}
    \end{center}
\end{figure}

Additionally, The ITk volume contains support structures, services, cooling systems and electronics accompanying the detector elements. The mechanical local support system is arranged in cylinders and rings interconnected to the global support. These are fabricated from low mass, high rigidity, high thermal conductivity materials which are carbon-based composites. The body of the Strips and Pixels is bounded by the ITk Outer Service Volume (OSV), shown in Figure~\ref{fig:Strips-OSV_XS}.
Radiation levels are predicted to exceed 17 MGy in the innermost Pixels region and 0.7 MGy in the Strips region, based on calculated fluences and applying a safety factor of 1.5~\cite{ATLAS-TDR-25}.

The focus of this work is a component of the Common Environmental Monitoring and Interlock System for the ITk, specifically the humidity monitoring.
The current ID is kept at a dew point of $-40^{\circ}$C, corresponding to a water content of 127 ppm; the ITk is planned to improve these values to a dew point of  at least $-60^{\circ}$C, corresponding to a water content of 10.5 ppm~\cite{ATLAS-TDR-25, ATLAS-TDR-30}.

There is always a possibility of humid air entering the ITk system, either via an air-leak or via diffusion through the flushing outlets or the service feed-throughs. There is a design overpressure of 10~mbar  to minimise this. The system is protected both in the case of an average operating temperature of $-20$~°C and a cooling fault which may lead to a temperature of  $-55$~°C.
The reason for monitoring dew point through relative humidity measurements is to prevent moisture condensation, which can damage electronics over time through corrosion and ice formation, potentially causing failure of detector components.
The design accommodates for this in several ways: dry nitrogen (N$_2$) flushing with a flushing gas dew point of $-80^{\circ}$C, corresponding to a water content of 0.54 ppm,  improved service feed-throughs, and humidity monitoring to detect the onset of leak events.  The ITk will reuse the current N$_2$ supply system. It will deploy novel fiber optic sensor (FOS) packages to provide relative humidity measurements. These FOS packages are radiation-resistant and designed for high sensitivity in the exceptionally dry ITk conditions, at relative humidity levels of RH $<10\%$~\cite{FOS1, FOS2}.

However, there are two problems. Firstly, the N$_2$ flushing may not be  sufficiently uniform over the entire ITk, leading to the formation of “dead zones” with low atmospheric renewal rates.
Secondly, the propagation of vapours from leak events to the sensors must be fast enough to provide sufficient time to mitigate condensation, requiring optimisation of the number and placement of sensors.
Thus, there is a need to understand the internal fluid environment of the ATLAS ITk in detail to determine areas of improvement in the flushing design that ensure dead zones with high humidity, low to no N$_2$ “coverage” and localised dew point rises outside the design range will not occur. In addition there is a need to determine the optimal number and placement of the humidity sensors.

Due to the complex thermoflow and transport phenomena, standard calculations and/or standards are insufficient to characterise the complex ITk volume. Computational Fluid Dynamics (CFD) has been utilized to address various engineering challenges~\cite{Verst,Patank,Ansorge,cui2014numerical,alizadeh2018numerical,palmowska2018research}. For example, it has been used to analyse the temperature, humidity, dew-point distributions and mass transfer in various engineering designs~\cite{yu2017high} including examining temperature and humidity distributions in an ice-rink~\cite{palmowska2018research}, modelling the temperature, humidity and dew-point in an evaporative cooler, modelling falling film heat and mass exchangers~\cite{heat-mass-exchangers} as well as in climate modelling such as quantifying the impact of extreme climate conditions on urban microclimates~\cite{micro-clim}. 

Therefore, in this work, CFD was used to develop a quantitative and qualitative understanding of the flow field, temperature, humidity and dew-point distributions in the ATLAS ITk as a result of the dry nitrogen purge under different leak conditions.
The idea is to understand the spatial region protected by a sensor, dead zones of low atmosphere renewal rates, and the propagation of vapour from leak events to sensors, under various operational and failure conditions.

A simplified model, which idealises the petals as discs and staves as rectangular plates was used.
Heat dissipation from the detector instruments and services is not modelled and the instruments are instead simplified as constant temperature boundaries. The model simulates humidity entering the volume by placing "leaks" at various points on the OSV, which act as inlets for humid air.
Despite these simplifications, the major elements responsible for accurate physics predictions were preserved. These simplifications are explained in more detail with justification in the following section.
The model allows for moist air leaks of different rates at specific points on  the OSV. Various parameters of interest were generated, including those related to the flow dynamics, the temperature, humidity and ultimately the dew point distributions.
The initial scope of work entailed modelling of the Strips region, which is presented in this study. However, subsequent work has begun on modelling the Strips with greater geometric detail, as well as the OSV and Pixel regions.



%% file: 2_Math_CFD.tex
\subsection{Simulation Geometry and Mesh}

\begin{figure}[!h]
    \begin{center}
        \includegraphics[width=0.70\linewidth]{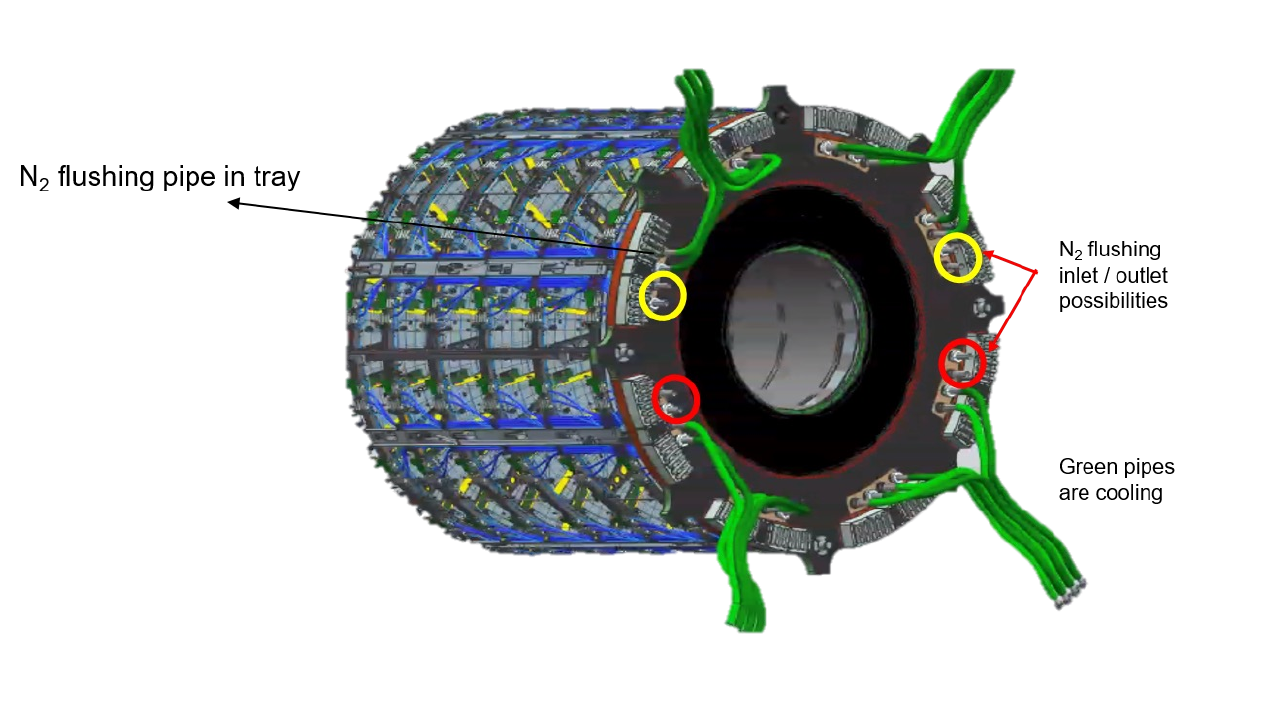}
        \caption{The initial CAD model of the Strips, with outlets, N$_2$ inlets and cooling pipes. \label{fig:flushing-CAD}}
    \end{center}
\end{figure}

\begin{figure}[!h]
  \begin{center}
    \subfloat[The CFD implementation of the manifold design
        \label{fig:manifold_a}]
        {\includegraphics[width=0.32\linewidth]{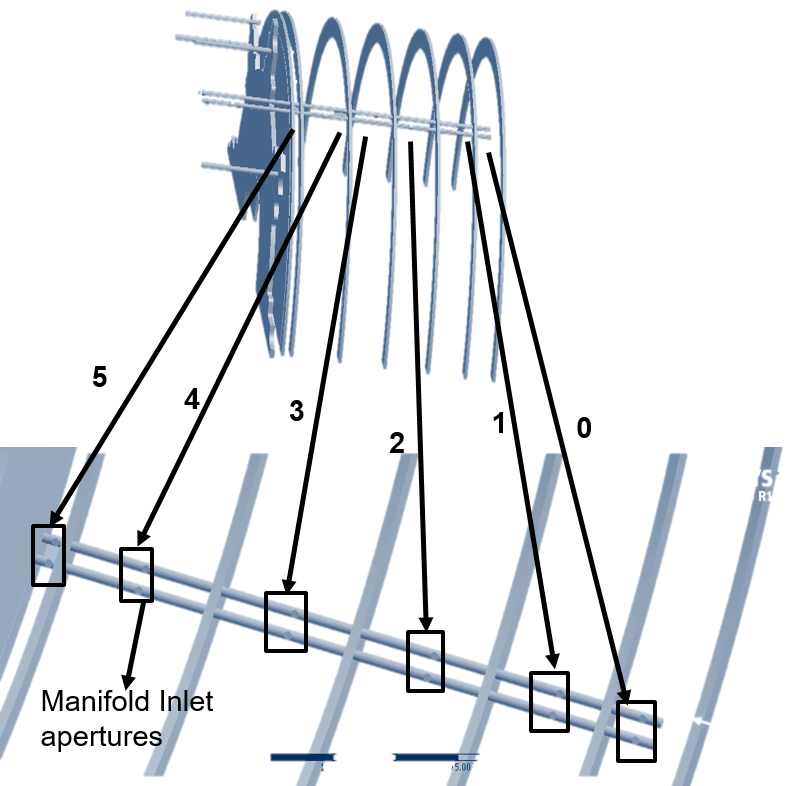}}
    \subfloat[The mechanical manifold design
        \label{fig:manifold_b}]
        {\includegraphics[width=0.62\linewidth]{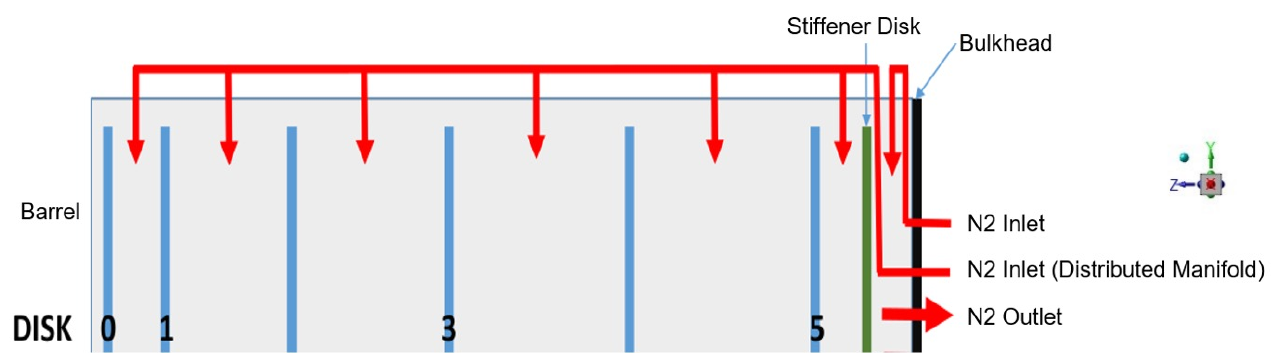}}
    \caption{The N$_2$ inlet manifold design using old inlet positions, showing (a) a 3D view of the manifold penetration through the bulkhead, with a zoom to show the inlet aperture locations, and (b) a schematic of the manifold design in the $yz$-plane.
    \label{fig:manifold}}
  \end{center}
\end{figure}

The ITk geometry is illustrated in Figures~\ref{fig:Strips-OSV_XS}~-~\ref{fig:manifold}.
The co-ordinate system used in this study is congruent with the ATLAS co-ordinate system; in Figure~\ref{fig:Strips-OSV_XS}, for example, the $x$-co-ordinate points towards to centre of the LHC ring (normal to the page), the $y$-co-ordinate points vertically upwards and the $z$-co-ordinate points along the beam pipe as shown in Figure~\ref{fig:ITk2}.
Due to the design of the ITk, as well as the position of the inlets and outlets, the model can leverage symmetry planes at $z=0~\si{m}$ and $x=0~\si{m}$. This allowed the size of the geometry to be simulated to be reduced by a quarter. Despite geometrical symmetry across the plane at $y=0~\si{m}$, a symmetry condition could not be applied to this plane due to the asymmetric effect of gravity in the negative-$y$ direction which would  violate the physics symmetry requirement.

\begin{figure}[!h]
     \begin{center}
        \includegraphics[width=0.84\linewidth]{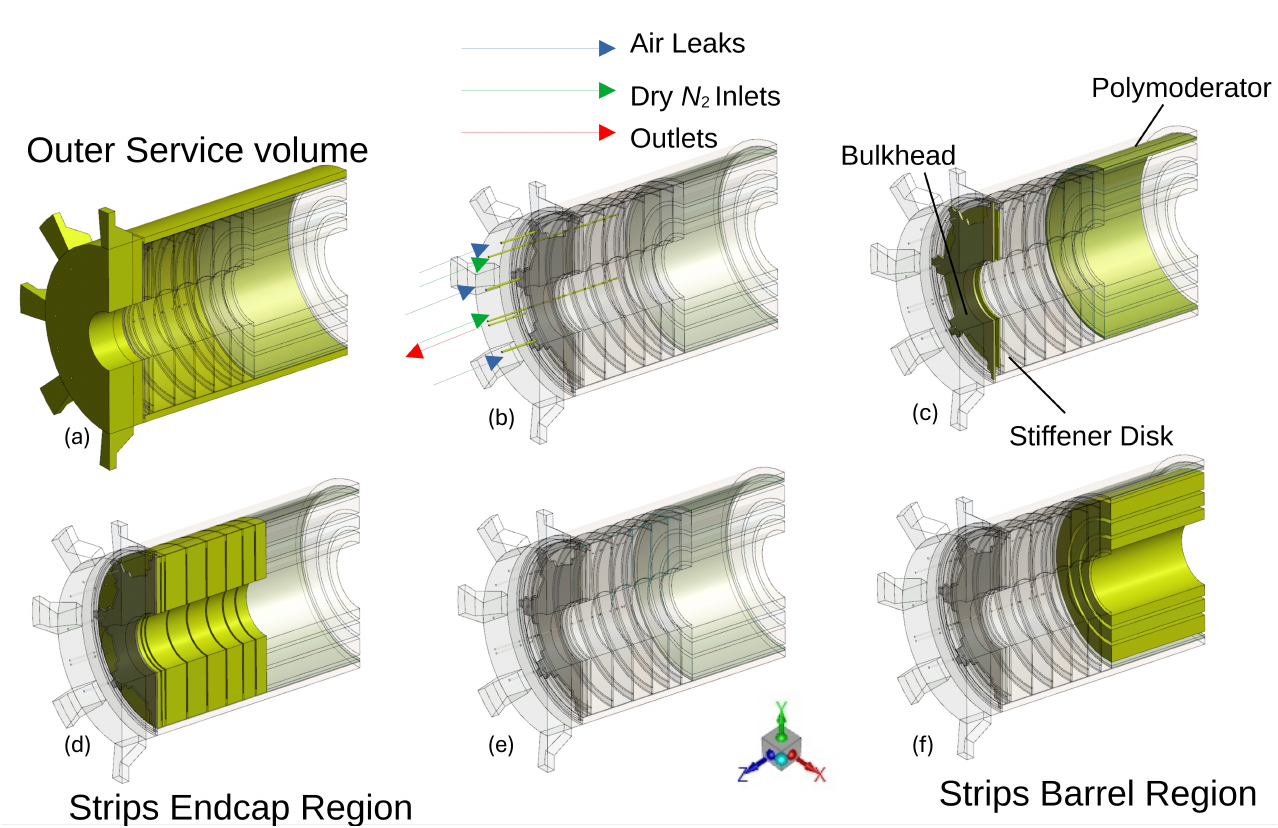}
        \caption{The ITk Geometry used in the CFD Model: (a) the fluid region of the OSV, (b) inlet, outlet and leak locations, (c) the solid components included in the model, (d) the fluid region of the Strips Endcap, (e) outlines of all solid components, and (f) the fluid region of the Strips Barrel.}
        \label{fig:detailed-geom}
     \end{center}
\end{figure}

The initial computer assisted design (CAD) design of the Strips Endcaps region is shown in Figure~\ref{fig:flushing-CAD}, with inlet locations circled in yellow while the outlets are circled in red. The green pipes are delivery mechanisms for the detector cooling systems; these are not modelled directly in this study, rather they are modelled as constant temperature boundary conditions, which is be discussed in further detail in Section~\ref{bound-cond}. Complex features from this design, such as wires and details on the detector disc surfaces, are abstracted away in the geometry used for simulations, shown in Figures~\ref{fig:Strips-OSV_XS} and~\ref{fig:detailed-geom}.

Figure~\ref{fig:Strips-OSV_XS} shows a simplified volume of the ITk in the $x=0$ plane. The green volume represents the ITk Strips region and the red volume indicates the OSV region, with inlet and outlet locations indicated by arrows. This model focuses on the ITk Strips, since the design of the flushing scheme for this region was a priority. The Pixels volume is not included in this model and is assumed to be thermally isolated from the Strips.

Figure~\ref{fig:detailed-geom} shows the quarter of the ITk volume which is simulated, with each panel highlighting a different feature of the geometry.
Panel (e) shows the outlines of the of the solid components - note that detector discs and support cylinders are modelled with flat, smooth surfaces.
The fluid regions are shown in panels (a), (d) and (f), indicating the OSV, Strips Endcaps and Strips Barrel respectively.
Panel (c) shows the solid components included in the simulations: the polymoderator, the Bulkhead and the stiffener disc. Other solid components, such as the detector discs and support cylinders, are represented in the simulations as constant boundary conditions, as discussed in Section~\ref{bound-cond}.
The OSV and Strips regions are modelled as separate fluid volumes, isolated from each other by the polymoderator in the Barrel region and the bulkhead head and "skin" in the Endcap region (the latter is too thin to display). Heat transfer between the two volumes occur and is modelled.
Lastly, panel (b) shows the locations of inlets, outlets and leaks.

The bulkhead and stiffener disc are structural members, and part of the mechanical local support system. The bulkhead specifically acts as a structural partition separating the OSV from the Strips Endcap. Whereas the stiffener disc acts as a structural support ring (member) to provide rigidity to the Endcap. The polymoderator is a neutron shielding component, specifically used to protect the sensors from damage induced by neutron back-scatter from the calorimeters~\cite{sahal}.

\begin{figure}[!ht]
    \begin{center}
        \subfloat[Old piping position
            \label{fig:inlets_a}]
            {\includegraphics[width=0.32\linewidth]{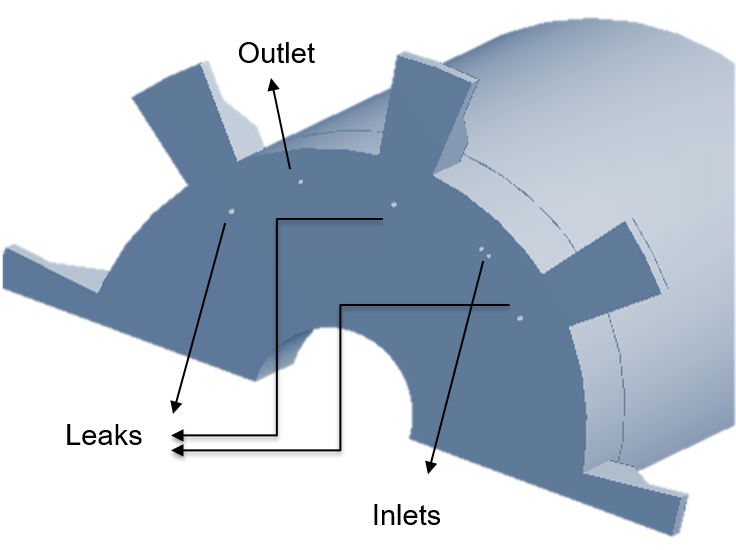}}
        \subfloat[New piping position
            \label{fig:inlets_b}]
            {\includegraphics[width=0.32\linewidth]{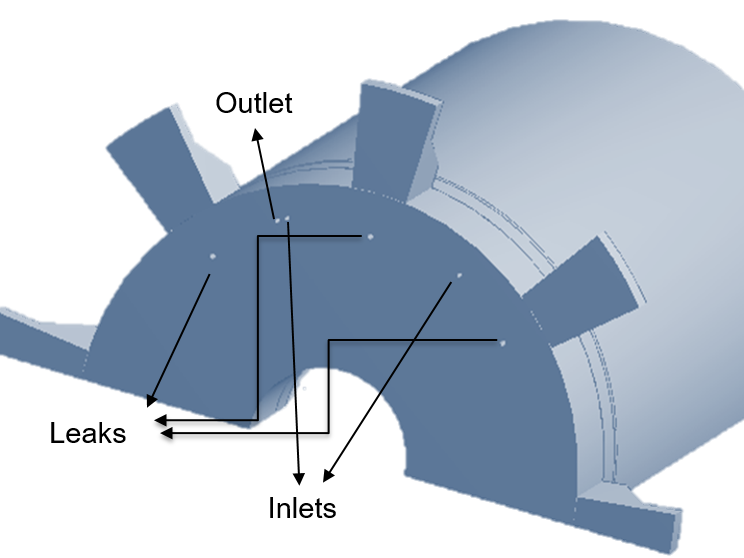}}
        \caption{The locations of the outlets, inlets and leaks on the OSV, showing (a) the old inlet positions and (b) the new inlet positions.
        \label{fig:inlets}}
    \end{center}
\end{figure}

Figure~\ref{fig:inlets} shows a quarter of the outer surface of the OSV, with one outlet, two inlets and three leaks. The position of each is repeated over the rest of the ITk volume so that there are four outlets, eight inlets and twelve leaks altogether. Based on initial simulations, the inlet positions were changed from the original locations in Figure~\ref{fig:inlets_a} to the new locations in Figure~\ref{fig:inlets_b}, where one inlet is moved close to the outlet.
This key change was made because the original design led to a notable stratification of temperature and humidity in the ITk, discussed further in Section~\ref{sec:no-leaks}.
The design also includes an inlet manifold that distributes N$_2$ between the stiffener disc and the disc shaped detector panel in the End Caps region. The final manifold aperture can also stream N$_2$ into the Barrel region. The layout of the manifold apertures (as per the original inlet positions) is displayed in Figure~\ref{fig:manifold}. 

A mixed mesh, specifically a mixture of structured hexahedral and un-structured tetrahedral cells, was used throughout the domain. Initially, the mesh cell count was 13 million. To ensure convergence and tractability of the simulation, the mesh was reduced to approximately 8.8 million cells whilst improving mesh quality. 
Poor quality cells were converted to polyhedral cells leading to a higher quality mesh. The base cell size was between $11~\si{mm}$ and $12~\si{mm}$. Due to limited computational resources, the small to large changes in geometry length scales and the complexity of the physics, a full mesh independence study was not possible. 
However, the mesh quality specifications of reference~\cite{Munoz} were achieved.

\subsection{Governing Equations}
The present work uses a CFD flow solver based on the finite volume method~\cite{Patank,Verst} to solve the Reynolds-Averaged Navier-Stokes equations and species transport. The flow is assumed to be steady state, incompressible flow. For the equations of state, the density is assumed to follow the ideal gas law and the dry nitrogen and humid air have temperature dependent viscosity. Thus, the conservation equations for continuity (Equation~\ref{eq1}) and momentum (Equation~\ref{eq2}) are represented as follows~\cite{flth}:
\begin{eqnarray}
    \label{eq1}
    \frac{\partial(u_i)}{\partial x_i}&=&0,\\
    	\label{eq2}
    \rho \frac{\partial (u_i u_j )}{\partial x_j}&=&-\frac{\partial p}{\partial x_i} + \frac{\partial} {\partial x_j}\left[ \mu \left( \frac{\partial \bar{u}_i}{\partial x_j} + \frac{\partial \bar{u}_j}{\partial x_i} - \frac{2}{3} \delta_{ij} \frac{\partial \bar{u}_l}{\partial x_l} \right)
    \right]-\rho\frac{\partial ( \overline{{u}^{\prime}_i {u}^{\prime}_j })}{ \partial x_j} -\rho g_i,
\end{eqnarray}
where $u_i$ and $\overline{{u}^{\prime}}_i$ represent the mean velocity and the mean turbulent fluctuating velocity, respectively. The fluid density is denoted by $\rho$, the molecular viscosity is $\mu$. The Reynolds stress tensor is represented by the term $\rho \overline{{u}^{\prime}_i {u}^{\prime}_j}= \tau_{ij}$.

The realizable $k$-$\varepsilon$ model is used for turbulence closure where the turbulent kinetic energy ($k$) and turbulent dissipation rate ($\varepsilon$) transport equations are represented as, respectively~\cite{flth}:
\begin{eqnarray}
    \label{eq4}
     \rho \frac{\partial (k u_{j})}{\partial x_j}
    &=&\frac{\partial}{\partial x_j}\left[\left(\mu +\frac{\mu_t }{\sigma_k}\right)\frac{\partial k}{\partial x_j}\right]+ G_k + G_b -\rho\varepsilon+ S_k,\\
    \label{eq5}
     \frac{\partial (\rho \varepsilon u_j)}{\partial x_j}
    &=&\frac{\partial}{\partial x_j}\left[\left(\mu +\frac{\mu_t }{\sigma_\varepsilon}\right)\frac{\partial \varepsilon}{\partial x_j}\right] + \rho C_{1}S\varepsilon - \rho C_{2}\frac{\varepsilon^{2}}{k+\sqrt{\nu \varepsilon}} + C_{1\varepsilon}\frac{\varepsilon}{k} C_{3\varepsilon}G_{b}+S_{\varepsilon}.
\end{eqnarray}
The turbulent viscosity $\mu_t$ is defined as $\mu_t=\rho C_{\mu} \frac{k^2}{\varepsilon}$. The $k$-$\varepsilon$ models use the Boussinesq hypothesis~\cite{Wilcox-turb, flth} to relate the Reynolds stress tensor $\tau_{ij}$ to the mean velocity gradients.
\begin{eqnarray}
    \tau_{ij}&=&-\rho \overline{{u}^{\prime}_i {u}^{\prime}_j }=\mu_t\left(\frac{\partial u_i }{\partial x_j} + \frac{\partial u_j }{\partial x_i}\right)-\frac{2}{3}\left(\rho k +\mu_t \frac{\partial u_k }{\partial x_k}\right) \delta_{ij}.
\end{eqnarray}
In addition, the transport and mixing of chemical species can be modelled by solving conservation equations that describe convection, diffusion, and reaction sources for each component~\cite{flth}. In the absence of chemical reactions, as is the case in this study, the species transport equation, for species $Y_i$ reduces to~\cite{flth}:
\begin{equation}
     \frac{\partial (\rho u_i Y_i)}{\partial x_j}=-\frac{\partial J_i}{\partial x_j} +S_{Y_i},
\end{equation}
where $Y_i$ is the local mass fraction of each species, $S_i$ is the rate of creation by addition from the dispersed phase plus any user-defined sources, $J_i$ is the diffusion flux of species $i$, which arises due to concentration gradients. Mass diffusion in turbulent flow $J_i$ is characterised by:
\begin{equation}
    J_i=-\left(\rho D_{i,m} + \frac{\mu_t}{S_{C_t}}\right)\frac{\partial Y_i}{\partial x_j} - D_{T,i} \frac{\frac{\partial T}{\partial x_{j}}}{T}
\end{equation}
where $D_{i,m}$ is the diffusion coefficient for species $i$ in the mixture, $ D_{T,i}$ is the thermal diffusion co-efficient for the species ${i}$ and $S_{c_t}$ is the turbulent Schmidt number which defined as
\begin{equation}
    S_{c_t}=\frac{\mu_t}{\rho D_t},
\end{equation}
where $\mu_t$ is the turbulent viscosity and $D_t$ is the turbulent diffusivity.

To model the heat transfer, the energy conservation equation can be expressed as follows:
\begin{eqnarray}
    u_{j}\frac{\partial (\rho E)}{\partial x_j} &=& -p\frac{\partial u_j}{\partial x_j} + k_t\frac{\partial^2 T}{\partial x_j \partial x_j} + \tau_{ij}\frac{\partial u_i}{\partial x_j} -\frac{\partial}{\partial x_j} \left(\sum_{j}^{n}m_j h_j \right)+ S_j,
    \label{energy-cons}
\end{eqnarray}
$k_t$ represents thermal conductivity, $T$ represents temperature, $E$ represents energy, and $S_j$ represents the source energy.

For the polymoderator, the bulkhead and the stiffener disc, Equation~\ref{energy-cons} simplifies to:
\begin{eqnarray}
    u_{j}\frac{\partial (\rho E)}{\partial x_j} &=& k_t\frac{\partial^2 T}{\partial x_j \partial x_j}
    \label{energy-cons-cond-3D}
\end{eqnarray}
Because the skin is approximately $3~\si{mm}$ thick, $3~\si{m}$ in length and over $2~\si{m}$ in diameter, the thin-walled assumption holds~\cite{flth}. Thus, the conduction term in the energy equation for the skin reduces to:
\begin{eqnarray}
    k_t\frac{\partial^2 T}{\partial x_j \partial x_j} \approx k_t\frac{\partial^2 T}{\partial n^2}
    \label{energy-cons-cond-1D}
\end{eqnarray}
where $n$ is the direction vector normal to the skin surface.

The dew point was calculated by implementing the Bögel modification of the Magnus formula~\cite{Bogel-Magnus}
\begin{equation}
    \label{dewp1}
    \gamma_m(T,R_h)=\ln\left(\frac{R_H}{100}e^{\left(b-\frac{T}{d}\right)\left(\frac{T}{c-T}\right)}\right),
\end{equation}
\begin{equation}
    \label{dewp2}
    T_{d_p}=\frac{c\gamma_m(T,R_h)}{b-\gamma_m(T,R_h)},
\end{equation}
where the constants $a=6.1121~\mbox{mbar}$, $b=18678~\mbox{mbar}$, $c=257.14^{\circ}$C, $d=234.5^{\circ}$C.
The accuracy of this correlation is $0.1\%$ in the temperature range of $-30^{\circ}C\leq T \leq-35 ^{\circ}$C.

\subsection{Solver Settings and Boundary Conditions}
\label{bound-cond}
To solve the system of equations proposed above, it was necessary to delimit the solution domain.
We started with a simplified ITk Strip geometry without leaks represented in Figure~\ref{fig:Strips-OSV_XS} as a starting point in the simulation process. 

\begin{table}[!ht]
    \centering
    \caption{Flow Boundary Conditions}
    \label{tab:flow-bc}
        \begin{tabular}{ l | l | l | l | l | l }
        \hline
        \textbf{Boundary} & \textbf{Type \textit{[unit]}} & \textbf{Value} & \textbf{Temperature} & \textbf{Species} & \textbf{Relative} \\
        & & & \textbf{[${^\circ}$C]} & & \textbf{Humidity $\%$} \\
        \hline
        Outlet & Pressure Outlet & 5 & Solver & Mixture & Solver \\
        (4 total) & \textit{[mbar]} & & Calculated & & Calculated \\
        \hline
        Inlets & Uniform Velocity & 2.4 & 15 & N$_2$ & 0 \\
        (8 manifolds) & Inlet \textit{[m/s]} & (per inlet) & & & \\
                      & Mass flow rate & 1.1 & & & \\
                      & Inlet \textit{[l/s]} & (total inlet) & & & \\
        \hline
        Leak rate 1 & Mass Flow & 0.1 (total) & $25$ & N$_2$, O$_2$, H$_{2}$O & 50-60 \\
        (12 total) & Inlet \textit{[l/s]} & 0.0083 (per leak) & & \\
        \hline
        Leak rate 2 & Mass Flow & 0.02 (total) & $25$ & N$_2$, O$_2$, H$_{2}$O & 50-60 \\
        (12 total) & Inlet \textit{[l/s]} & 0.0017 (per leak) & & \\
        \hline
        \end{tabular}
\end{table}

\begin{table}[ht!]
    \centering
    \caption{Wall Boundary Conditions}
    \label{tab:wall-bc}
        \begin{tabular}{ l | l | l | l | l }
            \hline
            \textbf{Boundary} & \textbf{Type} & \textbf{Heat Transfer Conditions} & \textbf{Temperature [$^{^\circ}$C]} & \textbf{Material} \\
            \hline
            OSV Outer Wall & Wall & Constant Temperature & 10 - 25 & Graphite \\
            \hline
             Barrel and Endcaps Detectors & Wall & Constant Temperature & $-25$ & Graphite \\
            \hline
            Bulkhead and Stiffener disc & Wall & Coupled & Fluent Calculated & Graphite \\
            \hline
            Polymoderator and Skin & Wall & Coupled & Fluent Calculated &  Glass \\
            \hline
            Inlet and outlet tubes & Wall & Coupled & Fluent Calculated &  PEEK \\
            \hline
        \end{tabular}
\end{table}

The gross (empty) ITK volume is 22~m$^3$~\cite{ATLAS-TDR-25, LACASTA2025170600} with a net (occupied) ITK volume of 13~m$^3$. The nominal flush rate by N$_2$ is 3900~l/hr,  corresponding to about  1.1~l/s or about  0.3 ITk vol/hr and a nominal overpressure of 4~mbar~\cite{Tomassini}.  
A maximum specification for the diffusion related leak rate from the cavern environment through the flushing outlets or the service feed-throughs is 10\% of the flush rate~\cite{ATLAS-TDR-25}.

The exact location of leaks in the ITk volume cannot be predicted and therefore, to study humidity, the model introduces fictitious leaks at locations in which they are most likely to occur. Leaks are placed at the PP1 feed-throughs near the bulkhead, shown in the top middle panel of Figure~\ref{fig:detailed-geom}. Two leak rates are investigated in this study, as shown in Table~\ref{tab:flow-bc}: $0.1~\mbox{l/s}$ and $0.02~\mbox{l/s}$. It was determined, during baseline calculations of design parameters and operational limits, that the maximum target leak rate is about 10$\%$ of the $N_{2}$ flush rate, which is $0.1~\mbox{l/s}$. Thus, the first case investigated the design performance at the maximum target leak rate. This was revised down by a factor of $5$ to $2\%$ of the $N_{2}$ flush rate, which is $0.02~\mbox{l/s}$, to determine the acceptable in-leak rate. Both leak rates introduce humid air at room temperature (25~°C) and the total leak rate is divided over the 12 leak points. The inlets pump N$_2$ at a velocity of 2.4~$\mbox{m/s}$. Humid air with three species was used: nitrogen, oxygen , and water vapour with respective mass fractions of 0.78, 0.21, and 0.01. Table~\ref{tab:flow-bc} and Table~\ref{tab:wall-bc} outline the boundary conditions. 

\begin{table}[ht!]
    \caption{Solver Settings}
    \begin{center}
        \begin{tabular}{l|>{\arraybackslash}p{0.7\linewidth}}
        \hline
            \textbf{Classification} & \textbf{Settings}  \\
        \hline
            \textsc{Solver}& - Pressure-Based 3D Coupled Solver (double precision)\\
                & - Steady state analysis\\
                & - Pressure Discretisation: Body Force Weighted \\
                & - Spatial Discretisation (all other transport equations): Second Order Upwind \\
        \hline
            \textsc{Energy equation}  & - Activated \\
        \hline
            \textsc{Viscous model}   & - Realizable $k-{\cal E}$ model\\
                & - Standards wall functions \\
                & - Full buoyancy effects \\
                & - Viscous heating effects \\
        \hline
            \textsc{Species model} & - Species transport \\
                & - Mixture properties \\
                & - Full Multicomponent Diffusion \\
        \hline
        \end{tabular}
    \end{center}
    \label{tab:Solver-BCs}
\end{table}

Not included in the tables is the temperature of the return coolant pipes (-40$^{\circ}$C). The main impact on fluid temperature is therefore the detector surfaces, which are assumed to be cooled to $-25$~°C and are modelled as boundaries of constant temperature, as shown in Table~\ref{tab:wall-bc}. Evaporative CO$_{2}$ cooling should dominate detector temperature, with convective heat exchange estimated to be a small perturbation. This should produce regions of varying temperatures on the detector surfaces, subsequently affecting local streaming of the gas. Such an effect might impact temperature stratification within the ITk volume.

Table~\ref{tab:Solver-BCs} lists the summary of the solver and model settings. The meaning of "coupled" for certain solid material components indicates that the heat flow through the solid will be modelled. In the case of the the Stiffener disc, the 3D Conduction method is used (Equation~\ref{energy-cons-cond-3D}). Thus, the stiffener disc was modelled as a solid region and meshed to compute the 3D conduction through the disc. It will be seen later that this allowed the temperature of the front and back of the stiffener disc to be computed. The same method was applied to the the bulkhead and Polymoderator. The skin is modelled using the 1D Fourier Conduction method as per Equation~\ref{energy-cons-cond-1D}. The simulation is run for three cases: first, ITk Strips without leaks; second, ITk Strips with leaks at a total leak rate of $0.1~\mbox{l/s}$ and last, ITk Strips with leaks at a total leak rate of $0.02~\mbox{l/s}$.

%% file: 3_results.tex
\subsection{ITk Strips model without leaks}
\label{sec:no-leaks}
\begin{figure}[!h]
    \begin{center}
        \subfloat[Pathline from inlet to outlet
            \label{fig:No-leaks-pathlines_a}]
            {\includegraphics[width=0.48\linewidth]{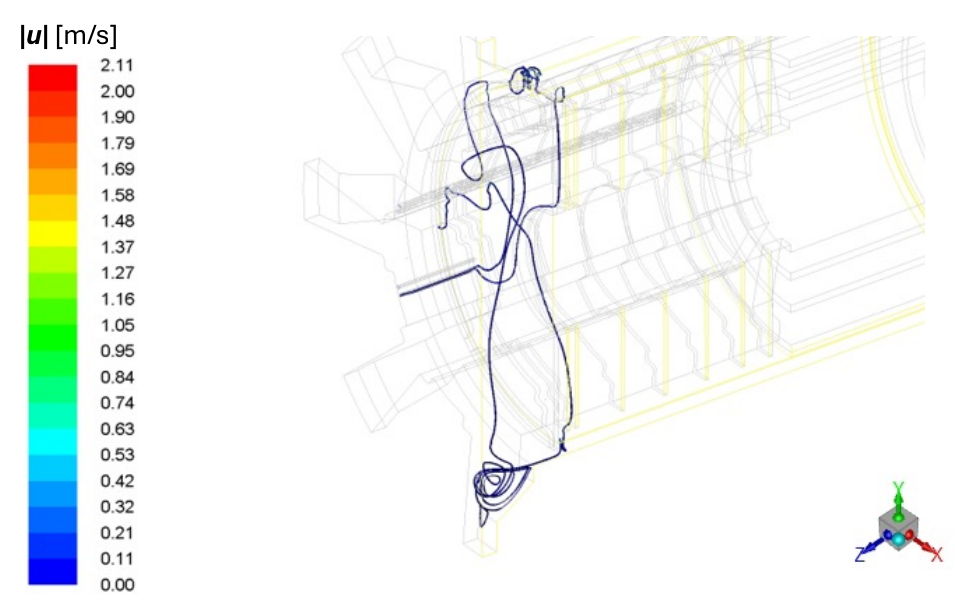}}
        \subfloat[Multiple inlet pathlines.
            \label{fig:No-leaks-pathlines_b}]
            {\includegraphics[width=0.42\linewidth]{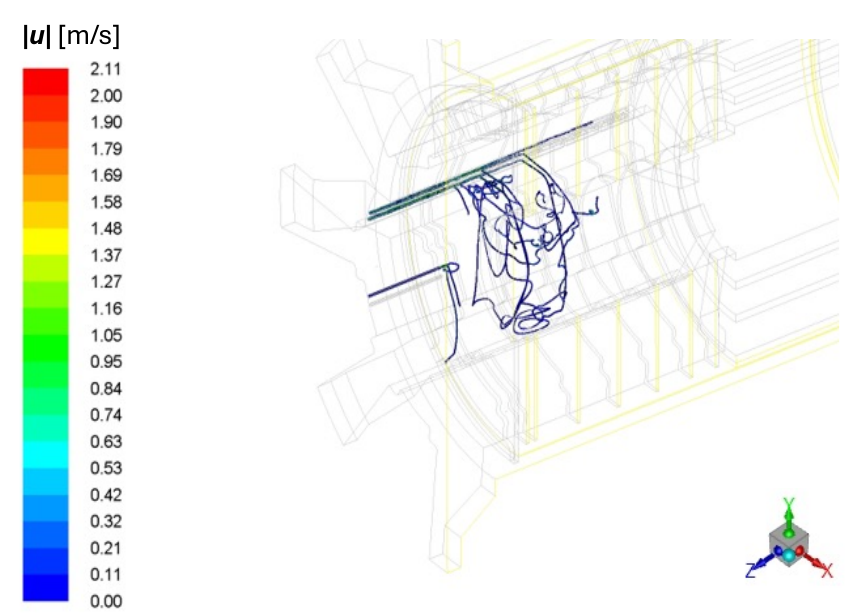}}
        \caption{The velocity pathlines [m/s] using the new piping positions - the average pathline velocity is around 0.1~m/s.
        \label{fig:No-leaks-pathlines}}
    \end{center}
\end{figure}

The CFD simulation results for ITk Strips without leaks  are considered as a limiting case where no humidity is introduced into the system.
Figure~\ref{fig:No-leaks-pathlines} displays the velocity pathlines from the inlets, using the new piping positions from Figure~\ref{fig:inlets_b}. The N$_2$ flow exhibits relatively low velocities, with pathlines remaining below 0.2 m/s. Given that the Strips volume extends 6~m along the $z$-axis, it is estimated that N$_2$ takes approximately 15 minutes to travel from the inlets to the centre.  As a result, the timescale for N$_2$ to disperse a localized buildup of humidity is estimated to be between 10 and 100 minutes.
A potential issue with the new piping positions is the possibility of short-circuiting the flow, as the inlet and outlet are positioned close together, allowing N$_2$ to travel only a short distance and potentially failing to flush the volume adequately. However, as shown in Figure~\ref{fig:No-leaks-pathlines_a}, this configuration effectively prevents short-circuiting, with N$_2$ traversing a significant portion of the volume before exiting.

Figure~\ref{fig:No-leaks-pathlines_b} reveals a that the inlet manifold from the new piping positions can lead to a non-uniform distribution of N$_2$ along the $z$-axis. This can be seen from the pathlines in the top half of the volume, which get shorter for nozzles closer to the centre of the Strips (lower $z$ values, in this case). This is due to decreased pressure in the nozzles further from the inlet and is a common problem with distributed manifolds. It is expected to result in decreased N$_2$ coverage for regions further from the inlets, and this problem can be addressed with different manifold designs in future simulations.

\subsection{ITk Strips model with leaks}
\begin{figure}[!h]
    \begin{center}
        \subfloat[Temperature 
            \label{fig:old-inlets-yz-plane_a}]
            {\includegraphics[width=0.42\linewidth]{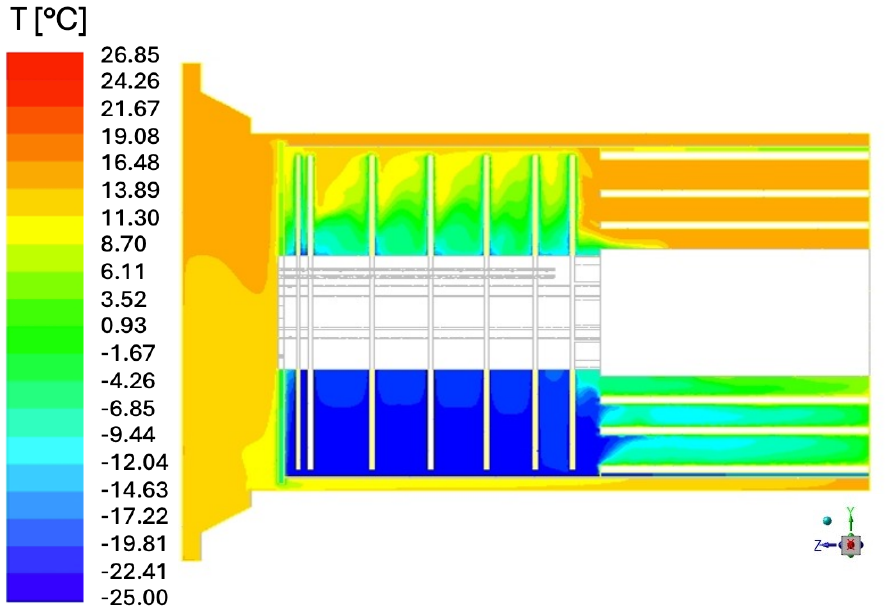}}
        \subfloat[Relative humidity 
            \label{fig:old-inlets-yz-plane_b}]
            {\includegraphics[width=0.37\linewidth]{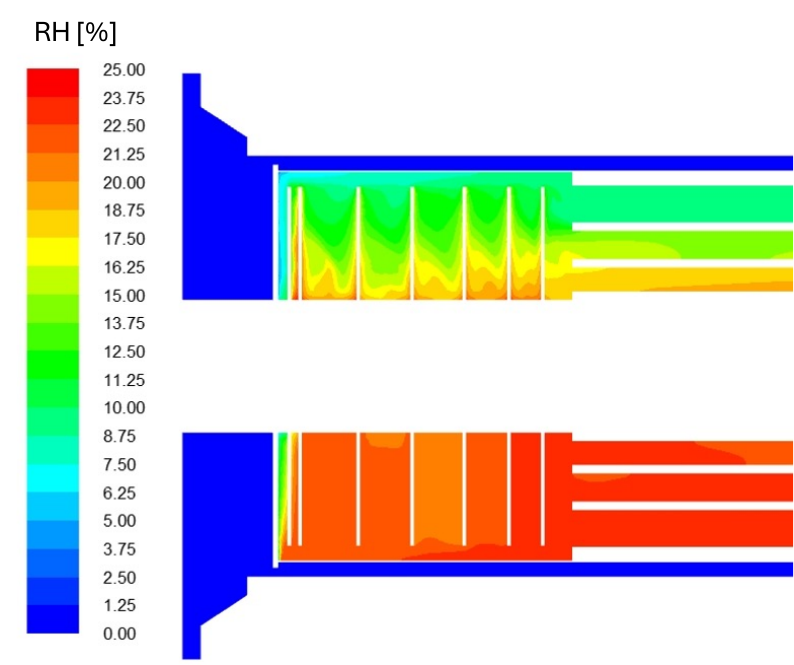}}
        \caption{(a) temperature [°C] and (b) relative humidity  [\%] in the $x=0$ plane using the old piping positions.
        \label{fig:old-inlets-yz-plane}}
    \end{center}
\end{figure}

\begin{figure}[!h]
    \begin{center}
        \subfloat[
            \label{fig:old-inlets-velocity_a}]
            {\includegraphics[width=0.32\linewidth]{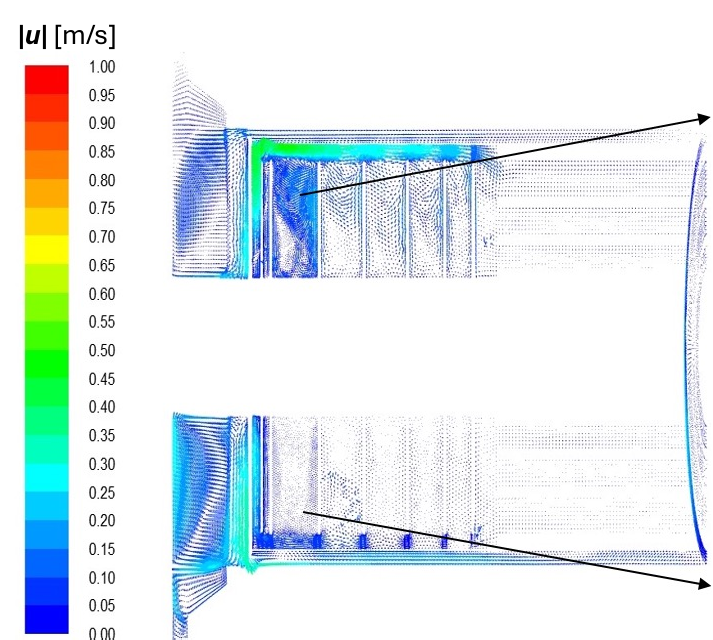}}
        \subfloat[
            \label{fig:old-inlets-velocity_b}]
            {\includegraphics[width=0.22\linewidth]{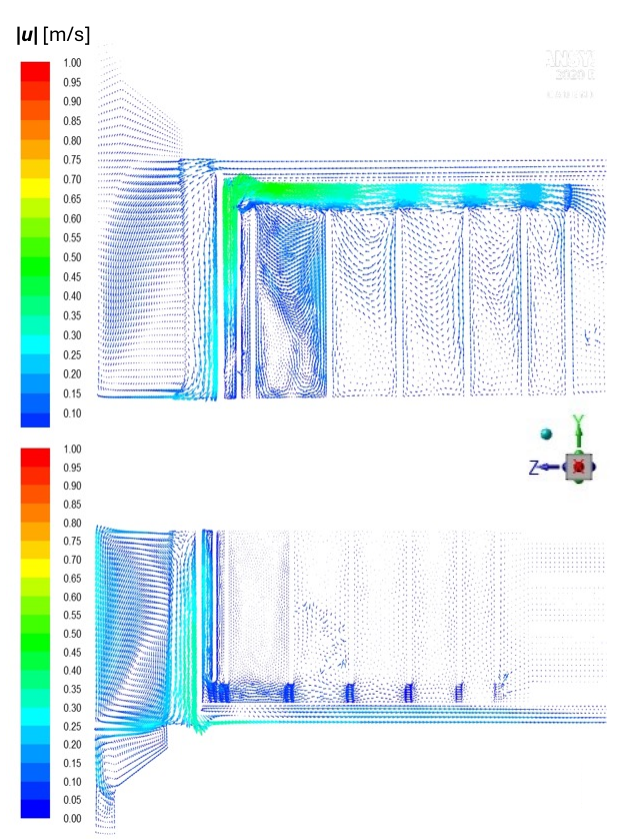}}
        \caption{Velocity vectors [$\mbox{m/s}$] in the $x=0$ plane using the old piping positions, showing (a) the full volume and (b) a zoomed-in view of the Endcaps.
        \label{fig:old-inlets-velocity}}
    \end{center}
\end{figure}

\begin{figure}[!h]
\centering
  \includegraphics[width=0.6\linewidth]{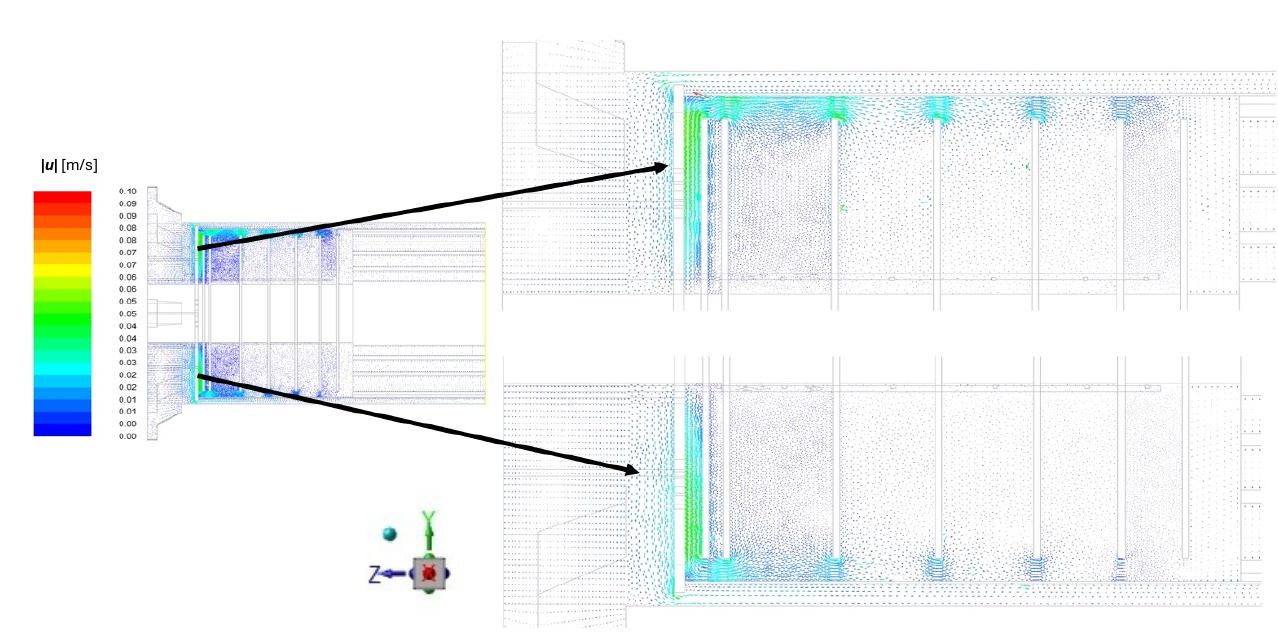}
  \caption{Velocity vector [$\mbox{m/s}$] in OSV, Strip Endcap and Strip Barrel in $x=0$ plane for the new piping position. 
  \label{fig:new-inlets-velocity}}
\end{figure}

\begin{figure}[!htb]
\begin{center}
  \includegraphics[width=0.65\linewidth]{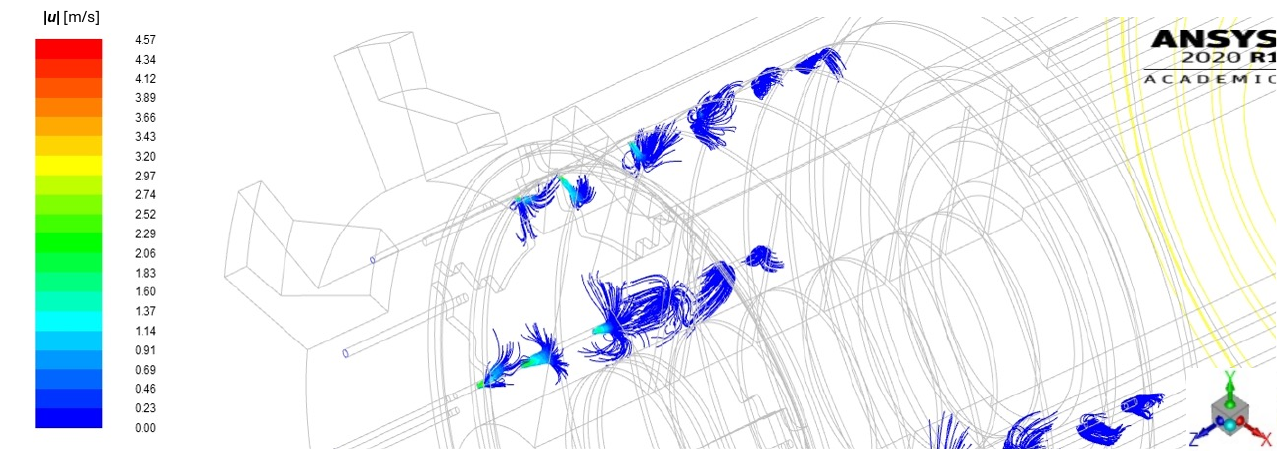}
  \caption{Velocity pathlines [$\mbox{m/s}$] released from the manifold nozzles. 
   \label{fig:new-inlets-velocity-manifolds}}
\end{center}
\end{figure}

\begin{figure}[!htb]
\begin{center}
  \includegraphics[width=0.70\linewidth]{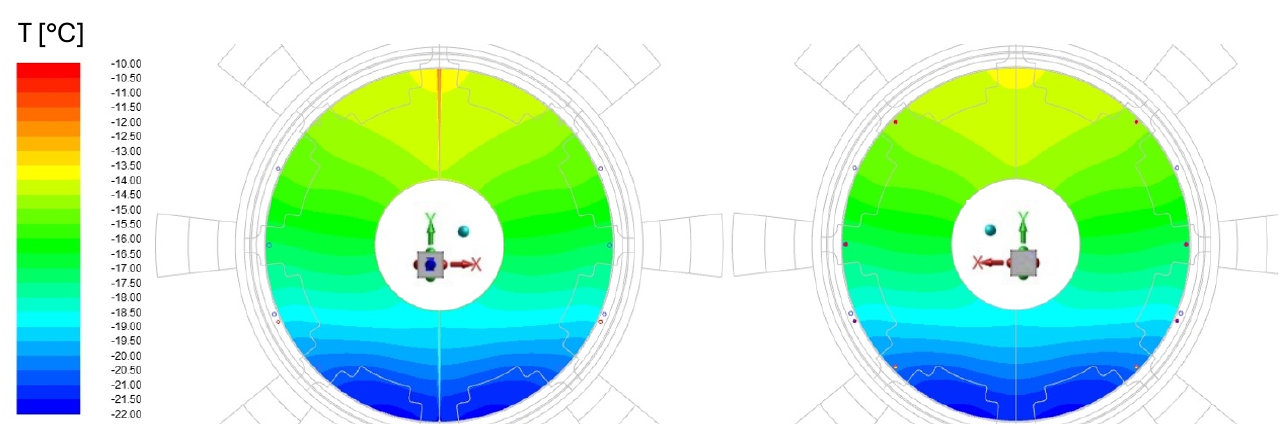}
  \caption{Temperature [°C] on the stiffener disc. The left figure is the bulkhead side, while the right is the Barrel side.\label{new-inlets-T-stiffener}}
\end{center}
\end{figure}

\begin{figure}[!h]
\begin{center}
  \includegraphics[width=0.75\linewidth]{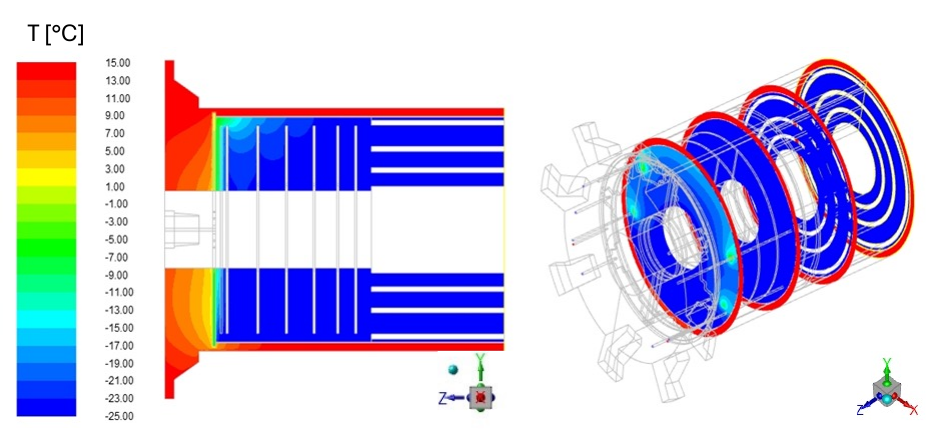}
  \caption{Temperature [°C] contours in (left) the $x=0$ plane and (right) four $xy$-plane slices along $z$ at the leak rate of $\mbox{0.1 l/s}$. 
  \label{fig:new-inlets-T_leak-rate1}}
\end{center}
\end{figure}

\begin{figure}[!h]
\begin{center}
  \includegraphics[width=0.75\linewidth]{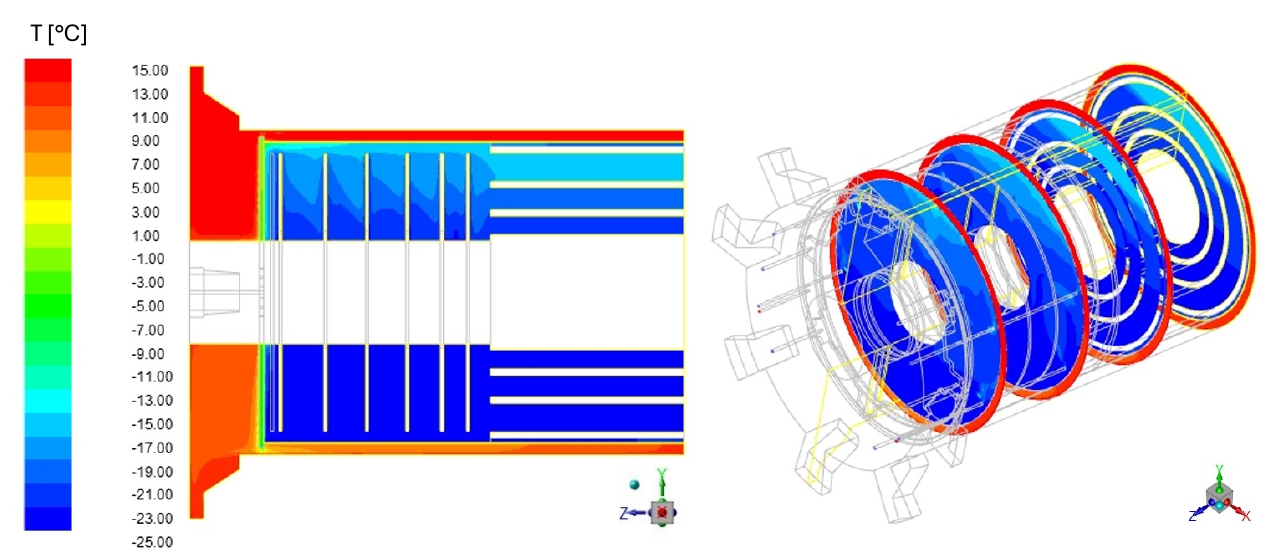}
  \caption{Temperature [°C] in (left) the $x=0$ plane and (right) four $xy$-plane slices along $z$ at the leak rate of $\mbox{0.02 l/s}$
  \label{fig:new-inlets-T_leak-rate2}}
\end{center}
\end{figure}

\begin{figure}[!h]
\begin{center}
  \includegraphics[width=0.75\linewidth]{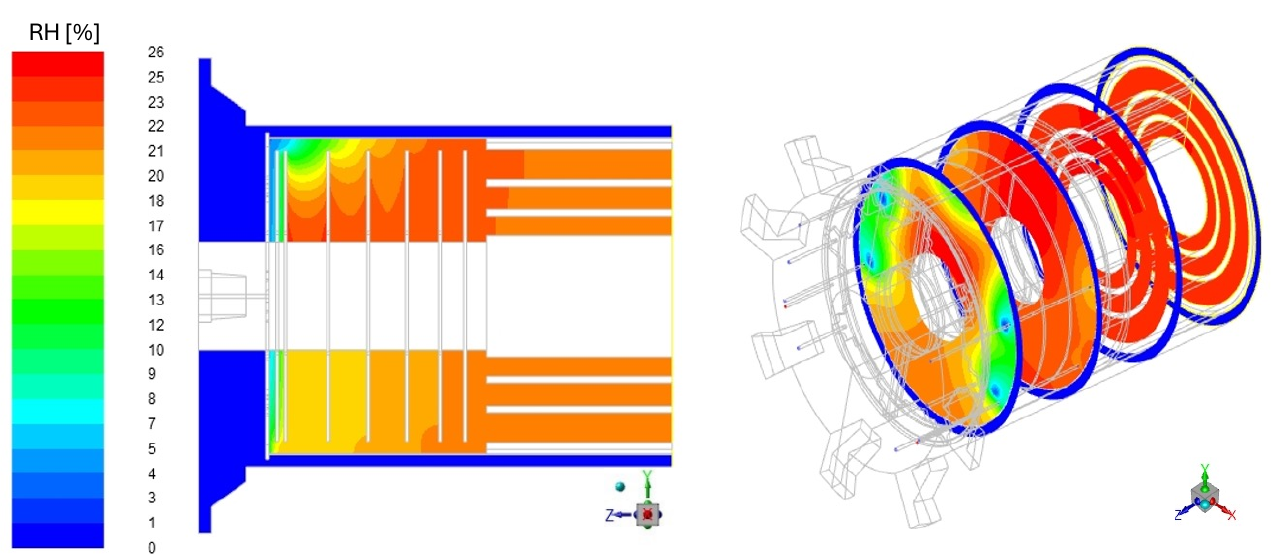}
  \caption{The relative humidity [$\%$] contours in (left) the $x=0$ plane and (right) four $xy$-plane slices along $z$ at the leak rate of $\mbox{0.1~l/s}$.
  \label{fig:new-inlets-RH_leak-rate1}}
  \end{center}
\end{figure}

\begin{figure}[!htb]
\begin{center}
  \includegraphics[width=0.75\linewidth]{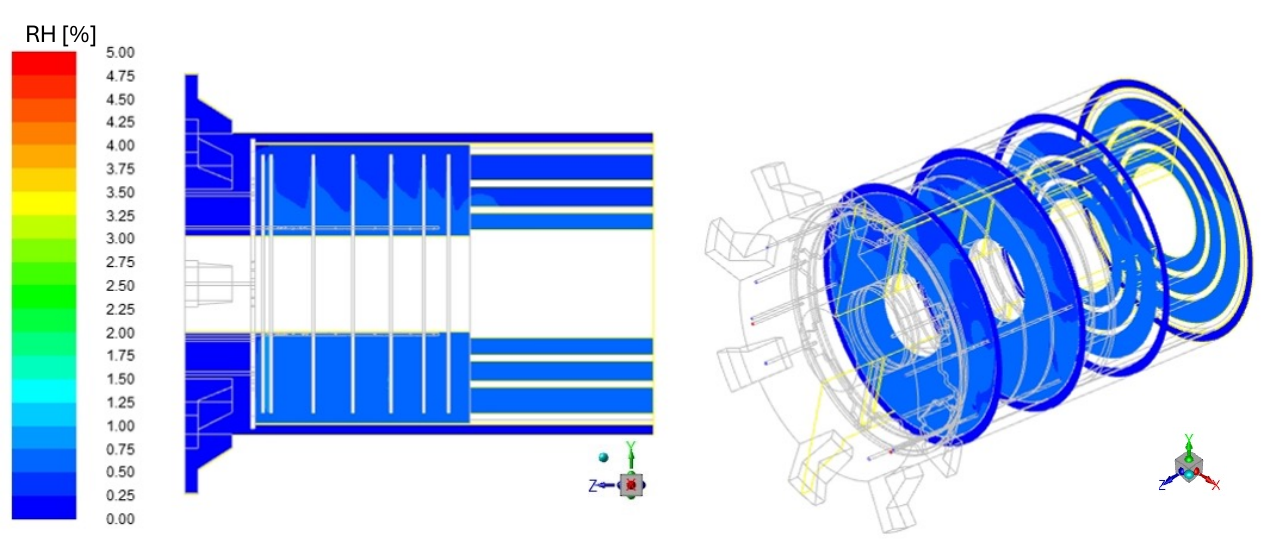}
  \caption{The relative humidity [$\%$] contours in (left) the $x=0$ plane and (right) four $xy$-plane slices along $z$ at the leak rate of $\mbox{0.02~l/s}$.  \label{fig:new-inlets-RH_leak-rate2}}
  \end{center}
\end{figure}

\begin{figure}[!ht]
\begin{center}
  \includegraphics[width=0.75\linewidth]{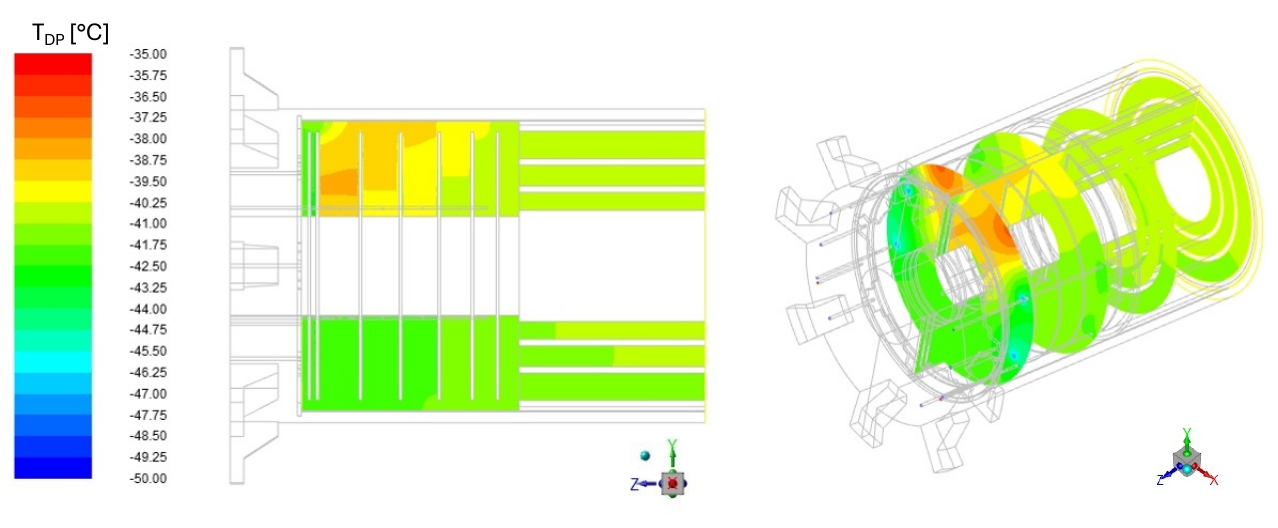}
  \caption{Local dew point temperature [$~^{\circ}$C] contours in (left) the $x=0$ plane and (right) four $xy$-plane slices along $z$ at a leak rate of  $\mbox{0.1l/s}$}
  \label{fig:new-inlets-DP_leak-rate1}
\end{center}
\end{figure}

\begin{figure}[!ht]
\begin{center}
  \includegraphics[width=0.75\linewidth]{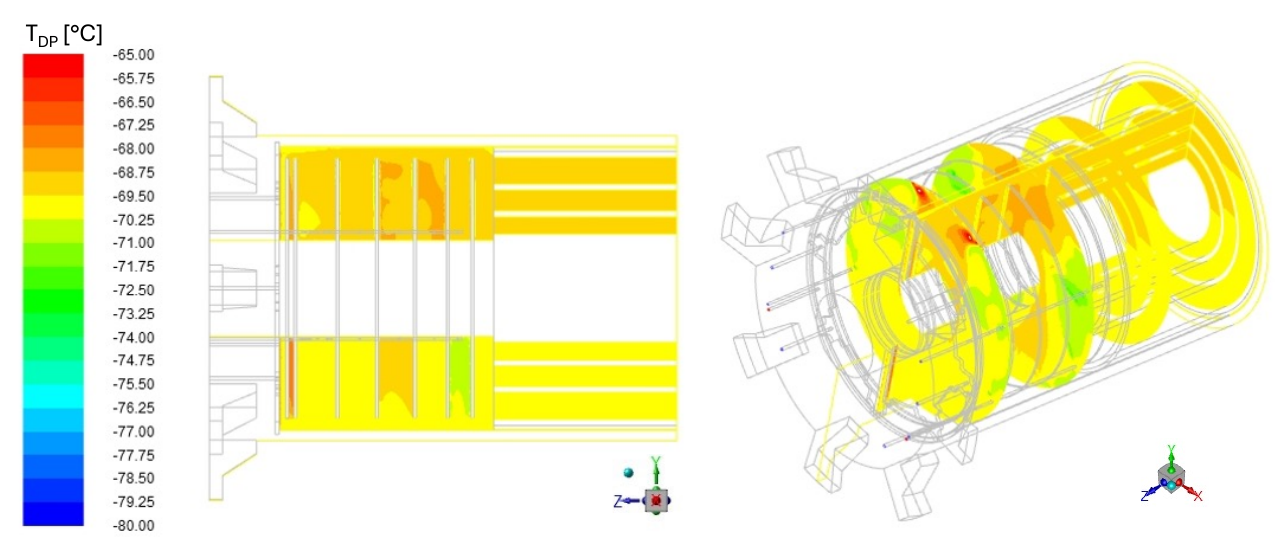}
  \caption{Local dew point temperature [$^{\circ}$C] contours in (left) the $x=0$ plane and (right) four $xy$-plane slices along $z$ at
  leak rate at 0.02 $\mbox{l/s}$}
  \label{fig:new-inlets-DP_leak-rate2}
	\end{center}
\end{figure}

The results of ITk Strips with leaks are shown in Figures~\ref{fig:old-inlets-velocity} to~\ref{fig:new-inlets-DP_leak-rate2}.
The velocity, temperature and relative humidity profiles for the old piping position (Figure~\ref{fig:inlets_a}) are shown in Figures~\ref{fig:old-inlets-velocity} and ~\ref{fig:old-inlets-yz-plane}. Note that the OSV has 0\% relative humidity throughout, as shown in Figure~\ref{fig:old-inlets-yz-plane_b}, but is not thermally isolated from the Strips, as shown in Figure~\ref{fig:old-inlets-yz-plane_a}. Each figure also shows a clear asymmetry between top and bottom (i.e. across the $y=0$ plane), particularly in the Strips Endcap. The Strips Endcap detectors are modelled by impermeable plane discs, forcing N$_2$ to pass between the OSV and the detector discs through the buoyancy effect.


Figure~\ref{fig:old-inlets-velocity} shows a nonuniform circulation of the flow and a low N$_2$ supply to the bottom of the Endcap and the Barrel regions, illustrated by the velocities and the higher density of vectors at the top of the volume.
It is observed in Figure~\ref{fig:old-inlets-yz-plane_a} that there is a significant temperature difference between the top and bottom of the Strips volume: around 25-35°C in the Endcaps and and 25°C in the Barrel region. This feature is driven by the buoyancy effect, causing the warmer (room temperature) N$_2$ entering the Strips region to move faster on the top region of the Endcap and Barrels.
This effect and the obstruction of the Endcap discs also leads to a non-uniform distribution of relative humidity, as shown in Figure~\ref{fig:old-inlets-yz-plane_b}. The relative humidity at the top of the Strips volume is around 10\%, compared to 25\% in the bottom. These temperature and humidity distributions would cause problems for detector performance and thus it was necessary to move the inlet pipe closer to the outlet, as shown in Figure.~\ref{fig:inlets_b}.

Moving to the new piping position, Figure~\ref{fig:new-inlets-velocity} reveals uniform circulation, a lower level of recirculation and and a high-speed region between the stiffener disc and the nearest detector disc. We observe a more uniform distribution from the new piping geometry than in the initial piping position. There is more N$_2$ supply at the bottom location of the Endcap Strips region. Also, high-velocity regions are found in the gap between the discs and the OSV's wall.
A non-uniform distribution with attenuating flow along the length of the manifold is observed in Figure~\ref{fig:new-inlets-velocity-manifolds}, as was observed in Figure~\ref{fig:No-leaks-pathlines_b}.

The temperature profile is presented in Figure~\ref{new-inlets-T-stiffener}. There is a higher temperature at the top of the stiffener disc compared to the bottom. The temperature gradient was calculated based on the average temperature on each side, and the calculation reveals that the gradient through the disc is $0.01^{\circ}$C. This value is less than the specified design specification value of $4^{\circ}$C, avoiding concerns of thermal deformation of the stiffener disc.


Temperature profiles in the $x=0$ plane as well as at a few chosen locations in the $z$-direction are shown in Figure~\ref{fig:new-inlets-T_leak-rate1} at a leak rate of $\mbox{0.1 l/s}$ and in Figure~\ref{fig:new-inlets-T_leak-rate2} at a leak rate of $\mbox{0.02 l/s}$. At both leak rates, there is a more uniform temperature distribution than in the initial piping position. It can be observed that the temperature is slightly higher in the OSV, bulkhead, stiffener disc, and first detector disc than in other places in the Endcap and Barrel regions.

The relative humidity distributions for both leak rates are displayed in Figures ~\ref{fig:new-inlets-RH_leak-rate1}~-~\ref{fig:new-inlets-RH_leak-rate2}, showing  zero humidity in the OSV as required and expected, as no leaks were modelled in the OSV. For the leak rate of $\mbox{0.02 l/s}$, we noticed a significantly lower relative humidity throughout the ITk volume compared to the leak rate of $\mbox{0.1 l/s}$. This profile can be explained by the change in the inlet's position, reducing the humidity in both Endcap and Barrel regions. Based on these results, the lower total leak rate of $\mbox{0.02 l/s}$ is the acceptable leak rate to stay within the design specification of a maximum $RH < 10\%$.

The dew point results for both leak rates are displayed below in Figures~\ref{fig:new-inlets-DP_leak-rate1} and~\ref{fig:new-inlets-DP_leak-rate2}.
As in the case of the temperature and humidity distributions, the dew point varies significantly close to the stiffener disc. Notable dew-point variations are observed higher up in the volume (in the $y$-direction). The volume average dew points are $-41.2~^{\circ}$C for the higher leak rate of $\mbox{0.1 l/s}$ and $-69.7~^{\circ}$C for the lower leak rate of $\mbox{0.02 l/s}$. The higher leak rate of $\mbox{0.1 l/s}$ exceeds the required specification of drier than $-60~^{\circ}$C. As expected, a lower leak rate of $\mbox{0.02 l/s}$ leads to a lower dew point, which is less than  the required specification of drier than $-60~^{\circ}$C. Based on these results, the lower total leak rate of $\mbox{0.02 l/s}$ is the acceptable leak rate to stay within the design specification of a maximum dew point that is within or below the design specification of drier than $-60~^{\circ}$C.

%% file: 4_discussion.tex
The aim of this work was to use CFD to develop a quantitative and qualitative understanding of the flow field, temperature, humidity and dew point distributions in the ATLAS ITk, specifically the ITk Strips region, as a result of the dry nitrogen flushing under different leak conditions. Additionally, initial results demonstrated that the intake manifold positioning in the original design would lead to non-uniform flow distribution with attenuation of the N$_2$ flow along the length of the Strips and in the lower region of the Strips. This would result in large areas with low to no atmospheric renewal of dry N$_2$, creating potential areas of high humidity or insufficient diffusion of the humid air and resultant high dew points leading to condensation on the detectors. These results prompted repositioning of the intake manifolds and subsequent results indicate that this new piping design improves N$_2$ coverage without introducing a short-circuit from inlets to outlets.

Based on the new design with the repositioned manifold, the model was used to assess the maximum leak rate that can be accommodated in case of an interlock failure or diffusion from the outlets due to the over-pressure. Based on the model results, the higher leak rate of $\mbox{0.1 l/s}$ is not acceptable and will lead to the dew point temperature exceeding the required specification of at or below $-60~^{\circ}$C. As expected, a lower leak rate of $\mbox{0.02 l/s}$ leads to a lower dew point, and this dew point is within the design specification of a dew point temperature at or below $-60~^{\circ}$C, thus, it is an acceptable leak rate.
Similarly, from these results, the lower total leak rate of $\mbox{0.02 l/s}$ is the acceptable leak rate to stay within the design specification of a maximum $RH < 10\%$. The findings indicate that, based on the latter leak rate and N$_2$ flushing rate, the ATLAS ITk Strips is expected to achieve a required level of "dryness". The model results provide an initial estimate of the maximum permissible leak rate to keep the dew point and relative humidity in the Strips Endcap and Barrel regions within the design specifications. The results also show that not only is this simulation useful in advising on humidity and dew point distributions in the ITk, but also on temperature distributions.

The new inlet positions presented in this study is a functional design that meets the design requirements and specifications. However, it is not an optimal design as the flow attenuates as the N$_2$ flows further downstream of the manifold, leading to sub-optimal flushing. Thus, a new manifold design is under investigation, whereby the nozzle sizes are varied along the length of the manifold with the intention of achieving uniform flushing across the the length of the manifold and through the ITk Strips. This manifold design and associated performance will be studied in future work.  

The temperature gradient across the stiffener disc, an essential structural support component, was calculated based on the average temperature on each side of the disc. The model results indicate that the temperature gradient through the disc is $0.01^{\circ}$C, which is less than the design specification value of $4^{\circ}$C, avoiding concerns of thermal deformation of the stiffener disc and ensuring the structural integrity of this critical support component.

It would be valuable to conduct a similar study that incorporates more realistic elements by replacing the discs and cylinders in the Strips Endcap and Barrel with petals and stave. This will allow us to understand the thermal flow behaviour with a higher degree of realism.
Future studies could also address the assumption of uniform temperature boundaries on the detector elements by allowing the temperature to vary over their surface. This can account for features such as hotspots due to certain electronic components and the effect of such on the temperature (and humidity) stratification in the ITk. Such changes are currently being applied in a follow up study.
The model predictions can only be validated against experimental data post commissioning of the ATLAS ITk Upgrade, scheduled for 2028-2030. Consequently the CFD will be used throughout the design cycle of the ITk Upgrade in line with the Conceive, Design, Implement and Operate (CDIO) engineering design paradigm.   


%% file: Acknowledgements.tex


We thank CERN for the very successful operation of the LHC, as well as the support staff from our institutions without whom ATLAS could not be operated efficiently.

The authors acknowledge the Centre for High Performance Computing (CHPC), South Africa, for providing computational resources to this research project


%% file: paperCFD1.bib
@book{2020-HL-LHC-TDR,
      author        = {Aberle, O. and Béjar Alonso, I and Brüning, O and Fessia, P and Rossi, L and Tavian, L and Zerlauth, M and Adorisio, C. and Adraktas, A. and Ady, M. and Albertone, J. and Alberty, L. and Alcaide Leon, M. and Alekou, A.},
      title         = "{High-Luminosity Large Hadron Collider (HL-LHC): Technical design report}",
      publisher     = "CERN",
      address       = "Geneva",
      series        = "CERN Yellow Reports: Monographs",
      year          = "2020",
      doi           = "10.23731/CYRM-2020-0010",
}

@Report{ATLAS-TDR-25,
    author         = "{ATLAS Collaboration}",
    title          = "{ATLAS Inner Tracker Strip Detector: Technical Design Report}",
    type           = "ATLAS-TDR-025; CERN-LHCC-2017-005",
    year           = "2017",
    note            = "\href{https://cds.cern.ch/record/2257755}{CERN Document Server}",
}

@Report{ATLAS-TDR-30,
    author         = "{ATLAS Collaboration}",
    title          = "{ATLAS Inner Tracker Pixel Detector: Technical Design Report}",
    type           = "ATLAS-TDR-030; CERN-LHCC-2017-021",
    year           = "2017",
    note            = "\href{https://cds.cern.ch/record/2285585}{CERN Document Server}",
}

@book{ATLAS:2019tdj,
    collaboration = "ATLAS",
    title = "{ATLAS}: {A 25-Year Insider Story of the LHC Experiment}",
    doi = "10.1142/11030",
    isbn = "978-981-327-179-1",
    publisher = "World Scientific",
    year = "2019"
}

@Book{Ansorge,
  author = 	 {R. Ansorge},
  title = 	{Mathematical Models of Fluid Dynamics: Modelling, Theory, Basic Numerical Facts - An Introduction},
  publisher = 	 {WILEY-VCH GmbH $\&$ Co. KGaA, Weinheim, Inc.},
  year = 	 {2003},
  address = 	 {Hamburg},
  edition = 	 {1},
}

@article{Bogel-Magnus,
      author = "Mark G. Lawrence",
      title = "The Relationship between relative humidity and the dewpoint temperature in moist Air: A simple conversion and applications",
      journal = "Bulletin of the American Meteorological Society",
      year = "2005",
      publisher = "American Meteorological Society",
      address = "Boston MA, USA",
      volume = "86",
      number = "2",
      doi = "10.1175/BAMS-86-2-225",
      pages=      "225 - 234",
}

@Article{FOS1,
    author         = {Consales, M. and others},
    title          = "{Nanoscale TiO2-coated LPGs as radiation- tolerant humidity sensors for high-energy physics applications}",
    journal        = "Opt. Lett.",
    volume         = "39",
    year           = "2014",
    pages          = "4128-4131",
    doi            = "",
    eprint         = "",
    archivePrefix  = "",
    primaryClass   = "",
}

@Article{FOS2,
    author         = {Berruti, G.},
    title          = "{Radiation tolerant fiber optic humidity sensors for High Energy Physics Applications}",
    journal        = "PhD Thesis, Universita’ degli Studi del Sannio",
    year           = "2015",
    pages          = "1",
    doi            = "",
    eprint         = "",
    archivePrefix  = "",
    primaryClass   = "",
}

@article{LACASTA2025170600,
title = {The ATLAS ITk strip detector system for the phase-II LHC upgrade},
journal = {Nuclear Instruments and Methods in Physics Research Section A: Accelerators, Spectrometers, Detectors and Associated Equipment},
volume = {1078},
pages = {170600},
year = {2025},
issn = {0168-9002},
doi = {https://doi.org/10.1016/j.nima.2025.170600},
author = {Carlos Lacasta}
}

@article{Munoz,
      author = "Gonzalo Anzaldo Munoz",
      title = "Lecture 3: Introduction to ANSYS Meshing",
      journal = "ANSYS Online Teaching",
      year = "2015",
      publisher = "ANSYS",
      address = " ",
      volume = " ",
      number = " ",
      pages=      "1 - 17",
}

@Article{PERF-2007-01,
    author         = "{ATLAS Collaboration}",
    title          = "{The ATLAS Experiment at the CERN Large Hadron Collider}",
    journal        = "JINST",
    volume         = "3",
    year           = "2008",
    pages          = "S08003",
    doi            = "10.1088/1748-0221/3/08/S08003",
    primaryClass   = "hep-ex",
}

@Book{Patank,
  author = 	 {S. V. Patankar},
  title = 	 "{Numerical Heat Transfer and Fluid Flow}",
  publisher = 	 {Hemisphere Publishing Corporation},
  year = 	 {1980},
  address = 	 {Minnesota},
  edition = 	 {1},
}

@techreport{Phase_2,
    author        = "ATLAS,  Collaboration",
    title         = "{Letter of Intent for the Phase-II Upgrade of the ATLAS Experiment}",
    institution   = "CERN",
    reportNumber  = "CERN-LHCC-2012-022, LHCC-I-023",
    address       = "Geneva",
    year          = "2012",
    note            = "\href{https://cds.cern.ch/record/1502664}{CERN Document Server}",
}

@article{Tomassini,
      author = "F.Rosatelli and S. Tomassini",
      title = "Leak Rate Specification for the Pixels PP1",
      journal = "ATLAS Technical Documentation AT2-IP-ES-0008",
      year = "2025",
      publisher = "ATLAS",
      address = " ",
      volume = " ",
      number = " ",
      pages=      "1 - 11",
}

@Book{Verst,
  author = 	 {K. H. Versteeg and W. Malalasekera},
  title = 	 {An Introduction to Computational Fluid Dynamics : The Finite Volume Method},
  publisher = {Longman Scientific and Technical},
  year = 	 {1995},
  address = {Essex},
  edition = {1},
}

@Book{Wilcox-turb,
  author = 	 {D. C. Wilcox},
  title = 	 "{Turbulence Modelling for CFD}",
  publisher = 	 {Griffin Printing, Inc.},
  year = 	 {1994},
  address = 	 {Glendale, CA},
  edition = 	 {2},
}

@article{alizadeh2018numerical,
  title={Numerical modeling and optimization of thermal comfort in building: Central composite design and CFD simulation},
  author={Alizadeh, M and Sadrameli, SM},
  journal={Energy and Buildings},
  volume={164},
  pages={187--202},
  year={2018},
  publisher={Elsevier}
}

@article{cui2014numerical,
  title={Numerical simulation of a novel energy-efficient dew-point evaporative air cooler},
  author={Cui, X and Chua, KJ and Yang, WM},
  journal={Applied energy},
  volume={136},
  pages={979--988},
  year={2014},
  publisher={Elsevier}
}

@Manual{flth,
  title = 	 {ANSYS Fluent 19.0 Theory Guide},
  author =   "{ANSYS Fluent Technical Staff}",
  organization = {ANSYS Inc.},
  address = 	 {Canonsburg, PA  USA},
  month = 	 {01},
  year = 	 {2019},
}

@article{heat-mass-exchangers,
title = {Fundamentals and applications of CFD technology on analyzing falling film heat and mass exchangers: A comprehensive review},
journal = {Applied Energy},
volume = {261},
pages = {114473},
year = {2020},
issn = {0306-2619},
doi = {https://doi.org/10.1016/j.apenergy.2019.114473},
author = {Tao Wen and Lin Lu and Weifeng He and Yunran Min},
}

@article{micro-clim,
title = {Validation of CFD models of urban microclimates under high temperature and humidity conditions during daytime heatwaves in dense low-rise areas},
journal = {Building and Environment},
volume = {266},
pages = {112087},
year = {2024},
issn = {0360-1323},
doi = {https://doi.org/10.1016/j.buildenv.2024.112087},
author = {Geunhan Kim and Gunwon Lee},
}

@article{palmowska2018research,
  title={Research on improving thermal and humidity conditions in a ventilated ice rink arena using a validated CFD model},
  author={Palmowska, Agnieszka and Lipska, Barbara},
  journal={International Journal of Refrigeration},
  volume={86},
  pages={373--387},
  year={2018},
  publisher={Elsevier}
}

@inproceedings{sahal,
  title={Simulation of the strip sub-detectors in the Inner Tracker of the ATLAS detector},
  author={Atkin, Ryna J and Yacoob, Sahal},
  booktitle={The Proceedings of SAIP2021, the 65th Annual Conference of the South African Institute of Physics, edited by Prof. Aletta Prinsloo, (UJ)},
  pages={193-198},
  year={2021},
  organization={South African Institute of Physics},
  isbn={ 978-0-620-97693-0}
}

@inproceedings{yu2017high,
  title={High Humidity Aerodynamic Effects Study on Offshore Wind Turbine Airfoil/Blade Performance Through CFD Analysis},
  author={Yu, Xue and Yan, Liu},
  booktitle={Turbo Expo: Power for Land, Sea, and Air},
  volume={50961},
  pages={V009T49A001},
  year={2017},
  organization={American Society of Mechanical Engineers}
}
